\documentclass[onecolumn,superscriptaddress,secnumarabic,amssymb, nobibnotes, aps]{revtex4-1}
\usepackage{amsmath,amssymb,graphicx, amsthm,wrapfig,epstopdf,esint,afterpage,marginnote,afterpage,lipsum,color,tabularx,ulem}
\usepackage{natbib}
\usepackage{enumitem}

\newcommand{\macro}[1]{\texttt{\textbackslash#1}}
\newcommand{\m}[1]{\macro{#1}}

\newcommand{\eps}{\epsilon}
\newcommand{\bra}[1]{\left(#1\right)}
\newcommand{\Bra}[1]{\left[#1\right]}

\begin{document}
\title{Controlling the interfacial and bulk concentrations of spontaneously charged colloids in non-polar media}%

\author{Sariel Bier}%
\affiliation{Department of Solar Energy and Environmental Physics, Blaustein Institutes for Desert Research (BIDR), Ben-Gurion University of the Negev, Sede Boqer Campus, Midreshet Ben-Gurion 8499000, Israel}

\author{Arik Yochelis}%
\email{yochelis@bgu.ac.il}
\affiliation{Department of Solar Energy and Environmental Physics, Blaustein Institutes for Desert Research (BIDR), Ben-Gurion University of the Negev, Sede Boqer Campus, Midreshet Ben-Gurion 8499000, Israel}
\affiliation{Department of Physics, Ben-Gurion University of the Negev, Be'er Sheva 8410501, Israel}
\date{\today}%

\begin{abstract}
	Stabilization and dispersion of electrical charge by colloids in non-polar media, such as nano-particles or inverse micelles, is significant for a variety of chemical and technological applications, ranging from drug delivery to e-ink. Many applications require knowledge about concentrations near the solid|liquid interface and the bulk, particularly in media where colloids exhibit spontaneous charging properties. By modification of the mean field equations to include the finite size effects that are typical in concentrated electrolytes along with disproportionation kinetics, and by considering high potentials, it is possible to evaluate the width of the condensed double layers near planar electrodes and the bulk concentrations of colloids at steady state. These quantities also provide an estimate of the minimum initial colloid concentration that is required to support electroneutrality in the dispersion bulk, and thus provide insights into the quasi-steady state currents that have been observed in inverse micellar media. 
\end{abstract}
	
\maketitle
	
\section{Introduction}

Spatiotemporal organization of electrical charges in nonpolar solvents, such as hydrocarbon solvents, is different due to the low dielectric constant of the solvent (which is typically smaller than 10) as compared to polar solvents {(e.g., water or alcohols)}, which are characterized by high-dielectric constants (about 80)~\cite{neyts2017}. {As a result, the Coulombic force in nonpolar solvents extends over larger distances and promotes the aggregation of nano-scale charged particles or the formation of solid precipitates~\cite{LYKLEMA2013116,DUKHIN201393,C8SM008D}. Nevertheless, colloids may be stabilized in a dispersion by the addition of surfactants and ligands, such as {the} macromolecular self-assembly of isolated inverse micelles, {which are} essentially a ternary water-in-oil system~\cite{smith2013controlling,prasad2016different}, and the references therein.} 

As opposed to salt-based electrolytes, in which the total number of charged species is usually conserved, experimental observations indicate that colloids in nonpolar solvents (inverse micelles or ligand coated nano-particles) can exchange charges via collisions, allowing neutral colloids to gain charge and charged colloids become neutral~\cite{Hsu2005}. The exchange of charge may occur via uneven distribution of encapsulated charged particles or impurities post-collision, or via the  transfer of charged surfactant molecules from one colloidal particle to another. The kinetic mechanism follows the disproportionation reaction~\cite{neyts2010,DUKHIN201393}, in which pairs of colliding neutral colloids exchange charges, yielding a pair of oppositely charged colloids, and vice versa: $p+n\rightleftharpoons m+m$, where $p,n$ and $m$ denote positive, negative and neutral inverse micelles, respectively. {Notably, while the dissociation kinetics of regular salts in aqueous electrolytes, $p+n \rightleftharpoons pn$, results in a fraction of charged colloids proportional to $\sqrt{[pn]}$, the disproportionation mechanism yields a charging fraction proportional to $[m]$~\cite{Hsu2005}. }

{Charge stabilization and conductivity of mono-dispersed colloids in nonpolar media have been advanced by incorporating disproportionation kinetics in the Poisson-Nernst-Planck (PNP) model equations, and employing blocking electrodes to separate the contributions of charge transfer reactions~\cite{neyts2006}. {In a series of works, in which voltage steps were applied to the electrodes and the subsequent depletion of charged colloids in the bulk was studied via transient currents, a qualitative description was obtained of the dependence of the amount of charge remaining in the bulk on various combinations of the system parameters, such as voltage, colloid concentration, and domain size,~\cite{neyts2017} and the references therein.} Specifically, the steady state in a system with a high initial concentration of charged colloids is different from that of a system with a low initial charge concentration; in the former there is a sufficient number of colloids for complete screening of the electrical field, i.e., to support elecroneutrality in the bulk. In the latter case, the low initial concentration of charge is insufficient for attaining bulk electroneutrality - all charges are attracted to the electrodes, the concentration of charged inverse micelles remaining in the bulk vanishes, and the electric field in the bulk remains finite. On the other hand, when neutral colloids are present, a parallel process occurs: as the neutral inverse micelles are driven toward the thermodynamic equilibrium ratio of neutral to charged inverse micelles, they undergo collisions to generate new charged colloids, which then drift towards the electrodes. If the process continues, the drifting colloids will contribute to a long-lived current, which resembles the currents observed in experiments~\cite{neyts2006}.
	
	The modified PNP equations, however, are valid only at low voltages and charge concentrations, since the finite size nature of the colloids is neglected. As a result, the accumulation of colloids at the electrode has no upper bound (maximal packing fraction) as the applied voltage is increased, i.e., the crowding effect is absent in this description. Therefore, in analogy with concentrated salt electrolytes, at high potentials we should expect to find an extended boundary layer near the electrodes, where the density attains its maximum value, and a bulk region around the center of the domain, where all densities are approximately homogeneous{~\cite{bikerman1942xxxix,borukhov1997steric,cervera2001ion,moreira2002simulations,cervera2003ion,kilic2007steric,biesheuvel2011two,hatlo2012electric,yochelis2014spatial,spruijt2014sedimentation}}. {The ability to estimate the width of this boundary layer or the density of the colloids remaining in the bulk could prove instrumental in the many applications involving stabilized charges at high potentials or high concentrations. These would include chemical processes in which surfactants are mixed with hydrocarbons, where the formation of inverse micelles affects the capacitance and insulation of the mixture, with important implications for performance and safety~\cite{morrison1993electrical}, or consumer devices such as e-ink displays, which rely on electrophoresis of charged pigments encapsulated within inverse micelles~\cite{comiskey1998electrophoretic}. The results may also allow greater control over systems that take advantage of microfluidics and droplet-based microreactors for chemical synthesis (e.g., as templates for nanoparticle fabrication~\cite{mori2001titanium,grzelczak2010directed,zhao2011nanoparticle} or for efficient synthesis of high-value chemicals such as pharmeceuticals~\cite{yoshida2011green,pileni1993reverse}), chemical analysis (such as lab-on-chip diagnostic devices~\cite{C2LC40630F}) or for drug and catalyst discovery~\cite{kreutz2010evolution,shang2017emerging}.}
	
	{Here we extend the mean field PNP-based model equations introduced in~\cite{neyts2006} to account for finite size effects, and derive analytic expressions that estimate the width of the double layer and the bulk densities. We provide a cutoff value for the amount of dilution by solvent above which complete screening of the electric field is unattainable due to the complete depletion of charge. We employ the framework introduced by Gavish-Elad-Yochelis~\cite{gavish_elad_yochelis}, since it captures several spatiotemporal properties of solutions containing mobile charges for a wide range of concentrations and particle sizes (including solvent-free~\cite{bguy}), and is therefore amenable to generalization~\cite{gavish_elad_yochelis}. The model equations thus provide a characterization of a medium in which the neutral part is comprised of two species: the reactive colloids, and the unreactive solvent. The analytical expressions are derived for the case in which the colloids are moderately diluted by solvent and a significant fraction of them are charged, and predict the critical value of dilution by the solvent, i.e, an estimation of the colloid concentration necessary for attaining electroneutrality. The results are shown to agree well with numerical integration.}
	
	\section{Model equations}
	
	We consider a medium comprised of monovalent, spherical and mono-dispersed positively charged colloids ($p$), negatively charged colloids ($n$), neutral colloids ($m$), and an inert solvent ($s$), {such that any volume element may be decomposed into the partial volume fractions of the components:}
	\begin{equation}\label{eq:steric}
	\frac{p}{p_{\rm{max}}}+\frac{n}{n_{\rm{max}}}+\frac{m}{m_{\rm{max}}}+\frac{s}{s_{\rm{max}}}=1,
	\end{equation}
	{with subscripts denoting the maximum concentration of each species.} Since the colloids are mono-dispersed, we define $\rho_{\rm max}\equiv p_{\rm{max}}=n_{\rm{max}}=m_{\rm{max}}$ and 
	\[
	\Upsilon \equiv \frac{s_{\rm{max}}}{\rho_{\rm max}}, 
	\] 
	as the relation between the packing density of the solvent and the colloids, respectively, such that $\Upsilon\gg 1$ corresponds to solvent molecules that are much smaller than the colloids, e.g., colloids in water.}

In addition, we make the following standard assumptions: (\textit{i}) Electric charges reside mostly within the colloids; the contribution of individual charged surfactant molecules in the bulk to the overall charge distribution is negligible, (\textit{ii}) the dielectric constant is homogeneous throughout the domain, and typically is taken as that of the solvent. While the dielectric constant may vary locally, e.g., in the vicinity of the inverse micelles due to the presence of the surfactant molecules, it is normally assumed that this radius of variation is sufficiently small to be ignored for realistic concentrations. (\textit{iii}) The interaction between colloids are summarized by the mass-action kinetics of the disproportionation model, where interactions such as bound pairs and aggregates of oppositely charged inverse micelles are assumed here to be energetically unfavorable transition states, (\textit{iv}) the ratio of charged colloids to neutral colloids may be large, even to the extent that the charged colloids outnumber the neutral ones. A physical realization of such a situation is found in the familiar electrophoretic display (e-ink) devices, where inverse micelles of opposite charges carry oppositely colored pigments in counter-propagating directions, with the applied voltage controlling the color that accumulates near the transparent electrode that functions as the screen~\cite{comiskey1998electrophoretic}.

The model starts with the free energy of the system:
\begin{align*}
\mathcal{F}=&\int {\rm dx} \Bigg\{ k_BT \bigg[ p\ln\bra{\frac{p}{p_i}}+n\ln\bra{\frac{n}{n_i}}+m\ln\bra{\frac{m}{m_i}}
+s\ln\bra{\frac{s}{s_i}} \bigg] + q(p-n)\phi-\frac{\eps}{2}|\nabla\phi|^2 \Bigg\}
\end{align*}
where the first term is the entropic contribution with $p_i=n_i$, $m_i$, $s_i$ being the respective initial concentrations, $k_B$ is the Boltzmann constant, and $T$ being the temperature, while the second term stands for the internal electrostatic energy, with $q$ being the elementary charge and $\eps$ is the dielectric permittivity (assumed here as constant). The relation between the initial concentrations of charged and neutral colloids is dictated by the disproportionation mass-action kinetic terms 
\begin{equation}\label{eq:disp_react}
\frac{{\rm d}p}{{\rm d} t}=\frac{{\rm d}n}{{\rm d} t}=-\frac{1}{2}\frac{{\rm d}m}{{\rm d} t}=-\alpha pn+\beta m^2, \\ 
\end{equation}
so that at the equilibrium $p_i=n_i$ and
\[
m_i\equiv\sqrt{{\alpha}/{\beta}}\cdot p_i=K p_i,
\]
where $\alpha$ and $\beta$ are forward and backward rates, respectively.

Following the Gavish-Elad-Yochelis framework~\cite{gavish_elad_yochelis} and the addition of charge generation~\cite{neyts2006}, the equations are given by 
{\begin{subequations}\label{eq:dimensional_model}
		\begin{equation} \label{eq:disMech}
		\frac{\partial}{\partial t}\left(\begin{array}{l}
		p\\n\\m\\ s
		\end{array}\right)= \nabla\cdot 
		\,\begin{bmatrix}
		\rho_{\rm max}-p & -p & - p& -p \\
		-n & \rho_{\rm max} -n &  -n & -n \\
		- m & - m & \rho_{\rm max}- m & -m\\
		-s & -s & -s & \Upsilon \rho_{\rm max}- s
		\end{bmatrix}
		\left(\begin{array}{l}
		p\, \mathbf{J}^p\\ 
		n\, \mathbf{J}^n\\
		m\, \mathbf{J}^m\\
		s\, \mathbf{J}^s
		\end{array} \right)
		\begin{array}{l}
		-\alpha pn+\beta m^2 \\ -\alpha pn+\beta m^2 \\+2\alpha pn-2\beta m^2 \\ \\
		\end{array}
		\end{equation}
		with the Coulombic interactions obeying the standard Poisson's equation
		\begin{equation}\label{eq:model_poss}
		\eps \nabla^2 \phi=q(n-p).
		\end{equation}
\end{subequations}}
The fluxes in~\eqref{eq:dimensional_model} are given by functional derivatives of the free energy
{\begin{align*}
	\mathbf{J}^{p}=\frac{\mu_p}{\rho_{\rm max}} \nabla \dfrac{\delta \mathcal{F}}{\delta p}, \quad \mathbf{J}^{n}=\frac{\mu_n}{\rho_{\rm max}} \nabla \dfrac{\delta \mathcal{F}}{\delta n}, \quad 
	\mathbf{J}^{m}=\frac{\mu_m}{\rho_{\rm max}} \nabla \dfrac{\delta \mathcal{F}}{\delta m}, \quad
	\mathbf{J}^{s}=\frac{\mu_s}{\rho_{\rm max}} \nabla \dfrac{\delta \mathcal{F}}{\delta s},
	\end{align*}}where $\mu$ is the mobility coefficient, assumed to be identical for all components, and which is related to the diffusion constant via the Einstein-Smoluchowski~\cite{islam2004einstein} relation $\mu= D/k_BT$, and $\delta$ denotes functional (variational) derivatives. {System~\eqref{eq:dimensional_model} is overdetermined, yet mass conservation allows elimination of the dynamic equation for the solvent via the relation $(p+n+m+s)_t=0$, so that the distribution for $s$ may be found at any time from $s=1-(p+n+m)/\Upsilon$, see~\eqref{eq:steric}. In the following discussion, we take the solvent molecule size to be equal to the colloid size, setting $\Upsilon=1$ (an appropriate assumption for large molecules, such as hydrocarbon solvents).}

In the absence of spatial symmetry breaking due to the formation of colloid aggregates~\cite{cheng1998properties,feldman2002non,ganguly2012investigating}, we focus our analysis on a one  dimensional spatial domain, such that $\nabla \to \partial_x$. We render the equations~\eqref{eq:dimensional_model} dimensionless by scaling the fields by $\rho_{\rm max}$ (see~\eqref{eq:steric}), the potential according to $\tilde{\phi}=q\phi/(k_BT)$ and the length by $\tilde{x}={x}/{\lambda}$, where $\lambda=\sqrt{{(\eps k_B T)}/{(q^2 \rho_{\rm max})}}$. Therefore, the effective domain size is $\tilde{L}\propto L\sqrt{\rho_{\rm max}}$, where, for a fixed real domain $L$, a large $\tilde{L}$ is equivalent to a large $\rho_{\rm max}$ (small colloids), and a small $\tilde{L}$ is equivalent to a small $\rho_{\rm max}$ (large colloids). Time is scaled as $\tilde t={t}/{\tau}$, where $\tau={\lambda^2}/{D}$, which leads to the dimensionless rate constants $k_1=\alpha \rho_{\rm max} \tau$ and $k_2=\beta \rho_{\rm max} \tau$ (with $\rho_{\rm max}$ having units of ${\text{mol}}/{\text{m}}$ and the units of $\alpha$ and $\beta$ being $\text{m}/\text{s}/ \text{mol}$). The dimensionless rate constants therefore satisfy the relation $\sqrt{k_1/k_2}=K$. Consequently, omitting the tildes, the dimensionless model equations read:
\begin{subequations}\label{eq:final_form}
	\begin{eqnarray}
	\partial_t p&=&\overbrace{\underbrace{\partial_x\left(p\partial_x\phi\right)+\partial^2_xp}_{\rm drift-diffusion}-\underbrace{\partial_x\left[p\bra{p-n}\partial_x\phi\right]}_{\rm finite~size}}^{\partial_x J_p} -\underbrace{k_1pn+k_2 m^2}_{\rm disproportionation}\\
	\partial_t n&=&\overbrace{\underbrace{-\partial_x\left(n\partial_x\phi\right)+\partial^2_xn}_{\rm drift-diffusion}-\underbrace{\partial_x\left[n\bra{p-n}\partial_x\phi\right]}_{\rm finite~size}}^{\partial_x J_n} -\underbrace{k_1pn+k_2 m^2}_{\rm disproportionation} \\
	\partial_t m&=&\overbrace{\underbrace{\partial^2_xm}_{\rm diffusion}-\underbrace{\partial_x\left[m(p-n)\partial_x\phi\right]}_{\rm finite~size}}^{\partial_x J_m} +\underbrace{2\bra{k_1 pn-k_2 m^2}}_{\rm disproportionation} \\
	&&\underbrace{\partial_x^2 \phi=n-p}_{\rm Poisson}
	\label{eq:poisson}
	\end{eqnarray}
\end{subequations}
{These equations differ from those proposed in~\cite{neyts2006} by the new finite-size terms}, which vanish together with one of the densities of charged colloids, i.e., when $(p,n)\to (1,0)$ or $(0,1)$, reflecting the fact that no drift can occur when there is complete depletion or complete saturation (close-packing) of charges. 

\section{Numerical solutions}
We first turn to numerical solutions of~\eqref{eq:final_form} to confirm both consistency with the transient currents obtained, for example, by Strubbe \textit{et al.}~\cite{neyts2006} as well as the realization of crowding near the domain boundaries at sufficiently high voltage. For this purpose we use no-flux (inert electrode) boundary conditions for all colloids 
\[
\left(\begin{array}{l}
J_p\\ 
J_n\\
J_m\\
\end{array} \right)\Bigg\vert_{x={\pm L/2}}=0,
\]
and Dirichlet (fixed voltage) for the potential 
\[
\phi|_{x=\pm L/2}=\pm V/2.
\]

From the initial electroneutral conditions $p_i=n_i$ and $m_i=Kp_i$, we define the total concentration of colloids (both charged and neutral) as
\[
c\equiv p_i+n_i+m_i=(2+K)p_i,
\]
so that $c\ll 1$ corresponds to a dilute solution containing mostly solvent, while $c\to 1$ corresponds to increasing the concentration of colloids. It is unclear however, whether high values of $c$ are actually attainable - above a certain density the inverse micelles can be expected to coalesce into layers~\cite{israelachvili2015intermolecular}, although {close packing of hard-sphere colloids cannot be totally excluded~\cite{moreira2002simulations}}. Additionally, {the dielectric permittivity at high concentrations may not be identical to that of the solvent~\cite{gavish2016dependence}}, although it has been shown elsewhere that the qualitative changes are negligible, even in cases where spatial symmetry breaking is present~\cite{gavish2017spatially}. In light of these limitations, we will limit $c$ to values below 0.8, which seems to be a reasonable upper bound at which the colloids may be expected to exist as separate entities. By rewriting the initial densities of the colloid species in terms of total concentration and reactivity, we obtain the initial species concentrations $p_i=n_i={c}/{(2+K)}$, and $m_i={Kc}/{(2+K)}$. The ratio of charged to neutral colloids at equilibrium is $2/K$, so for $K<2$ there are more charged colloids than neutral. The initial solvent concentration obeys $s_i=1-c$. 

\begin{figure}[tp]
	(a)\includegraphics[width=0.5\textwidth]{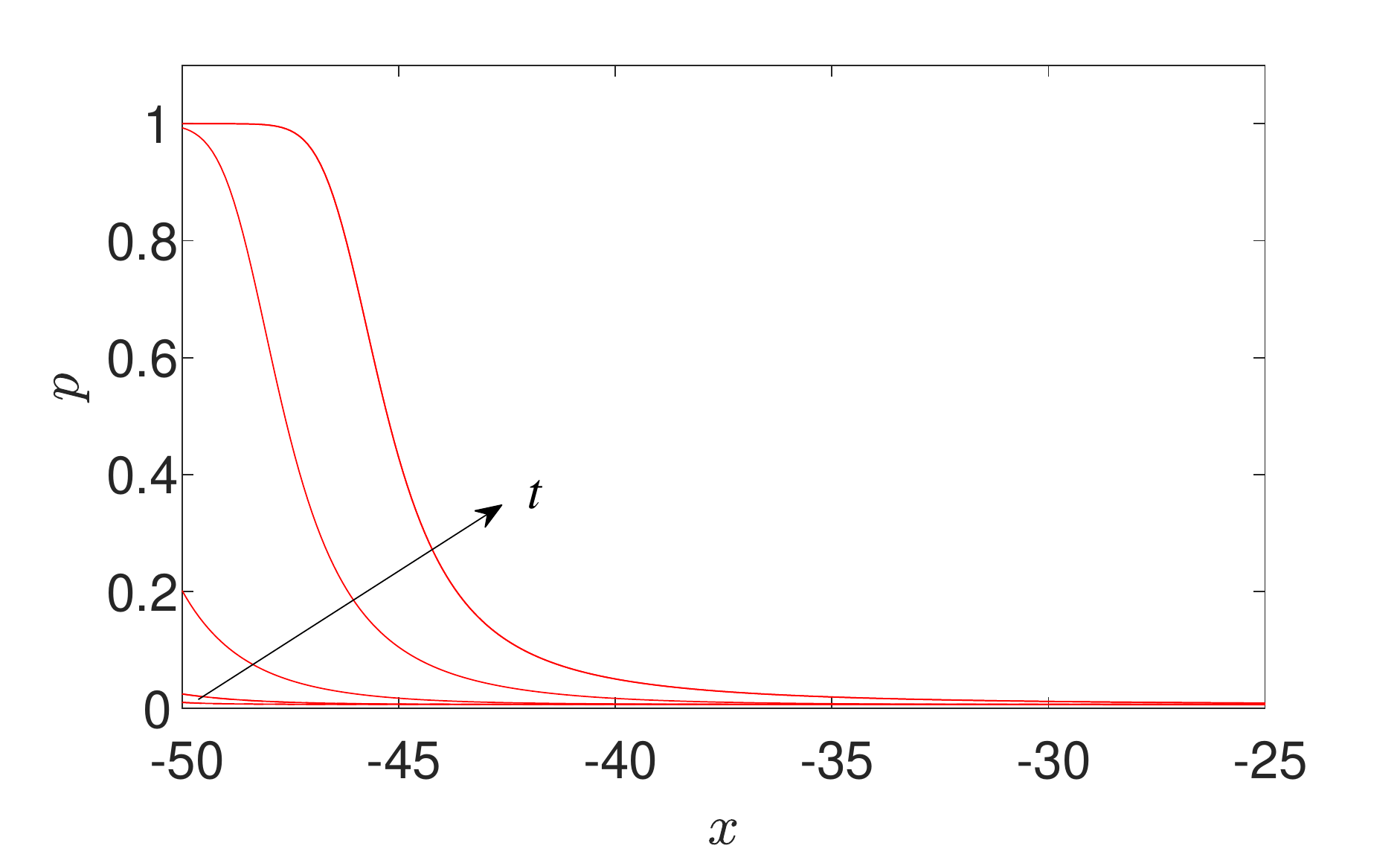}
	(b)\includegraphics[width=0.5\textwidth]{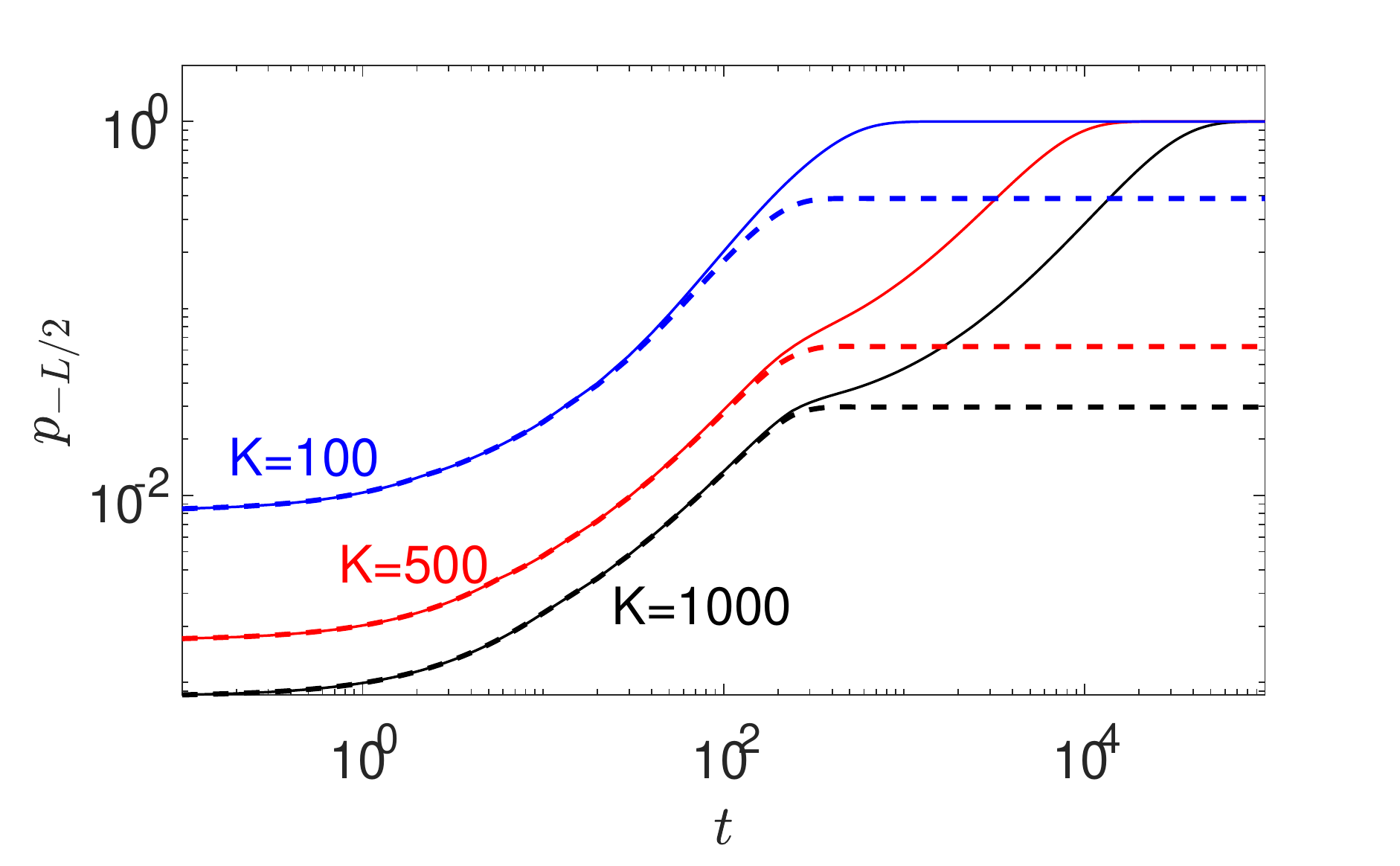}
	(c)\includegraphics[width=0.5\textwidth]{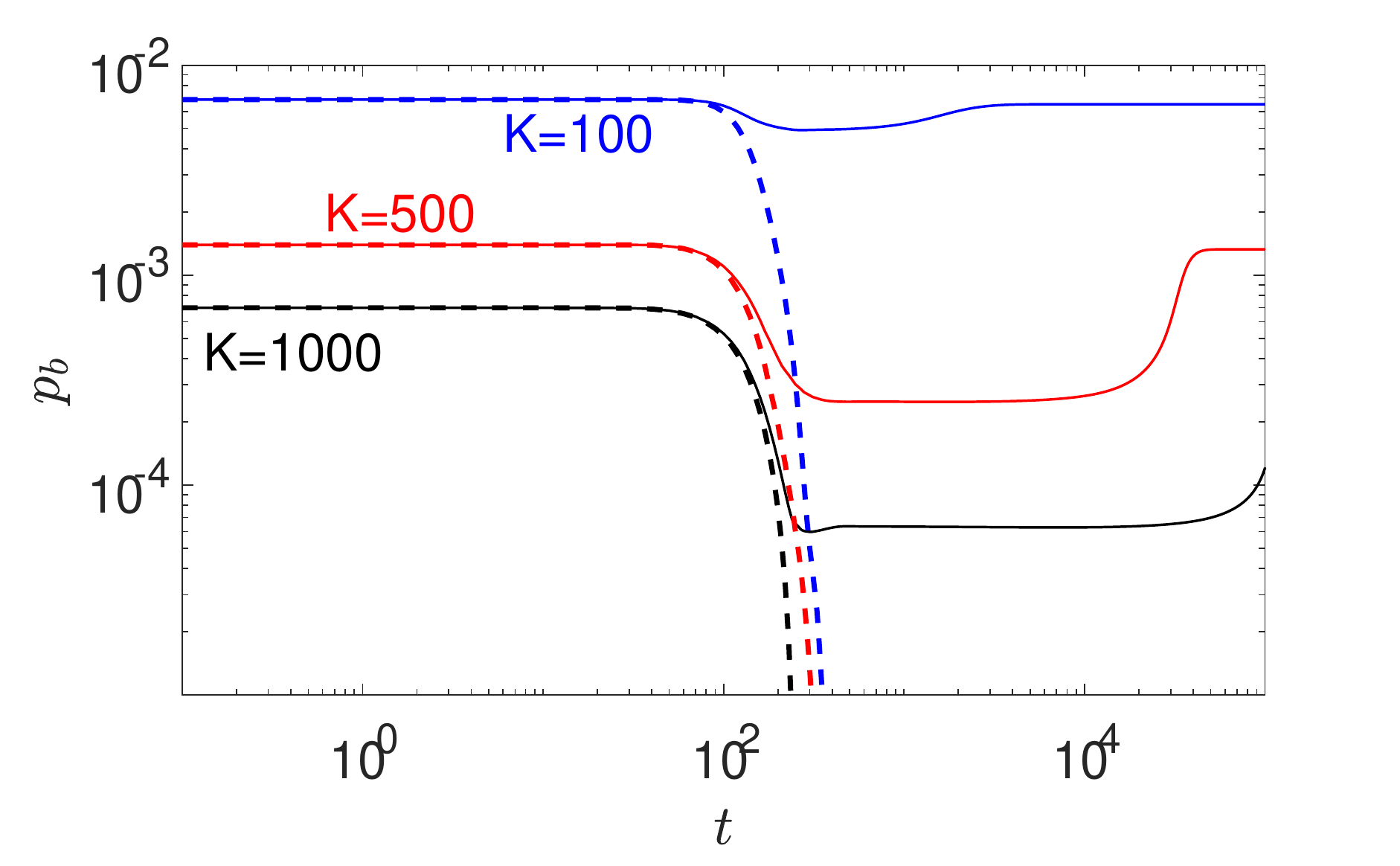}
	\caption{(a) Profiles of the positive charge $p$, taken at $t=1, 10, 10^2, 10^3, 10^4, 10^5$ for $K=100$, $V=40$, $c=0.7$ and $L=100$. (b) The change in $p$ at the boundary (electrode surface) ($x=-L/2$) over time, with $V$, $L$ and $c$ the same as above, and values of $K$ as labeled. Dashed lines depict the behavior without disproportionation reactions~\eqref{eq:disp_react}, while solid lines show the behavior when reactions are included. (c) The change of $p$ over time in the bulk ($x=0$), which shows complete bulk depletion without reactions (dashed lines) and partial depletion with reactions (solid lines). Higher values of $K$ correspond to a larger initial concentration of neutral micelles, and therefore a higher degree of dilution, allowing for comparison of cases with and without reactions. For the values of $K$ shown, higher dilution leads to a faster depletion of bulk charge and lower asymptotic densities at the electrode when reactions are absent, yet when reactions are included maximal densities of $p=1$ are reached and a residual charge dictated by the equilibrium constant $K$ remains.} \label{fig:p_prof}
\end{figure}

\begin{figure}[tp]
	(a)\includegraphics[width=0.5\textwidth]{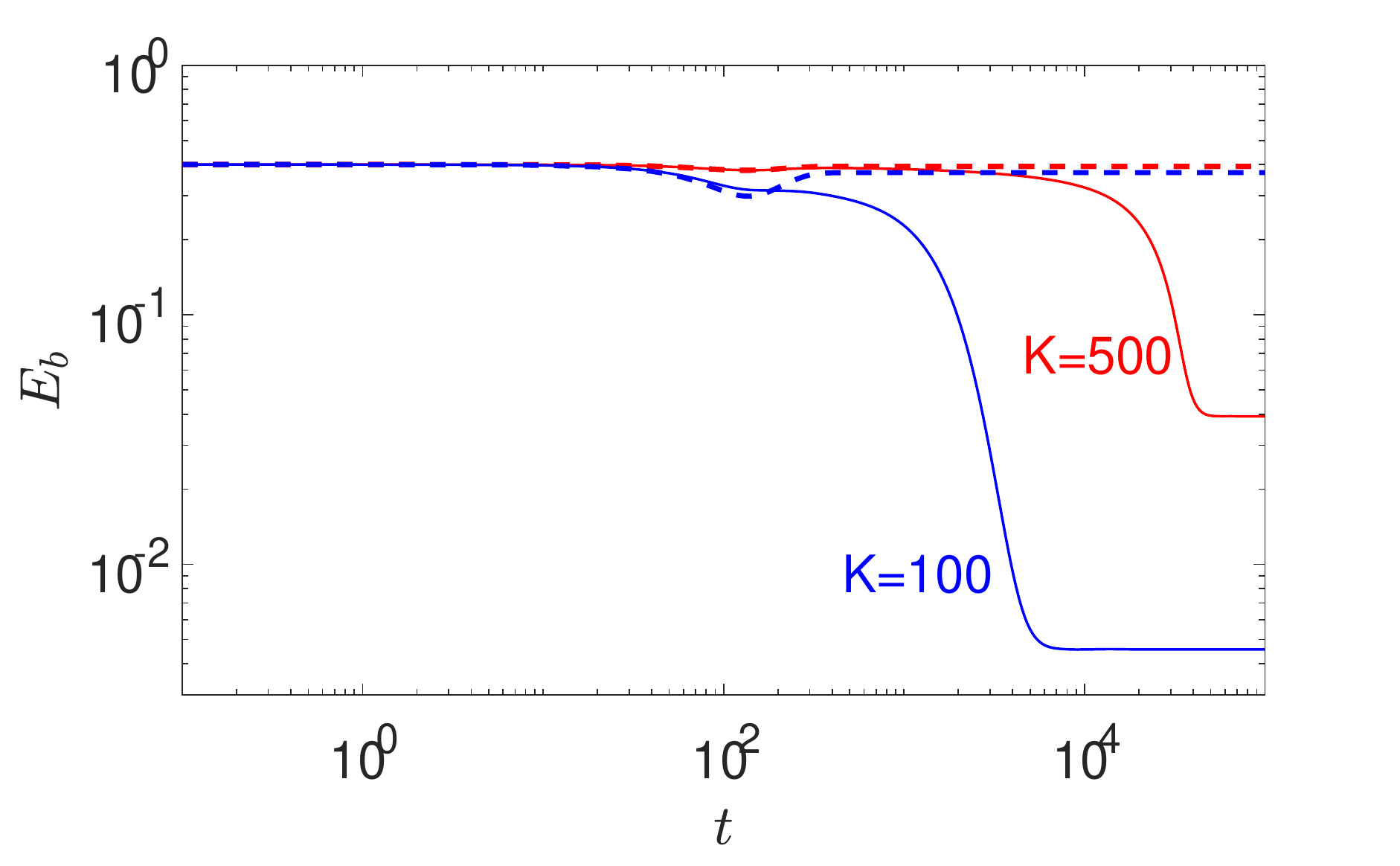}
	(b)\includegraphics[width=0.5\textwidth]{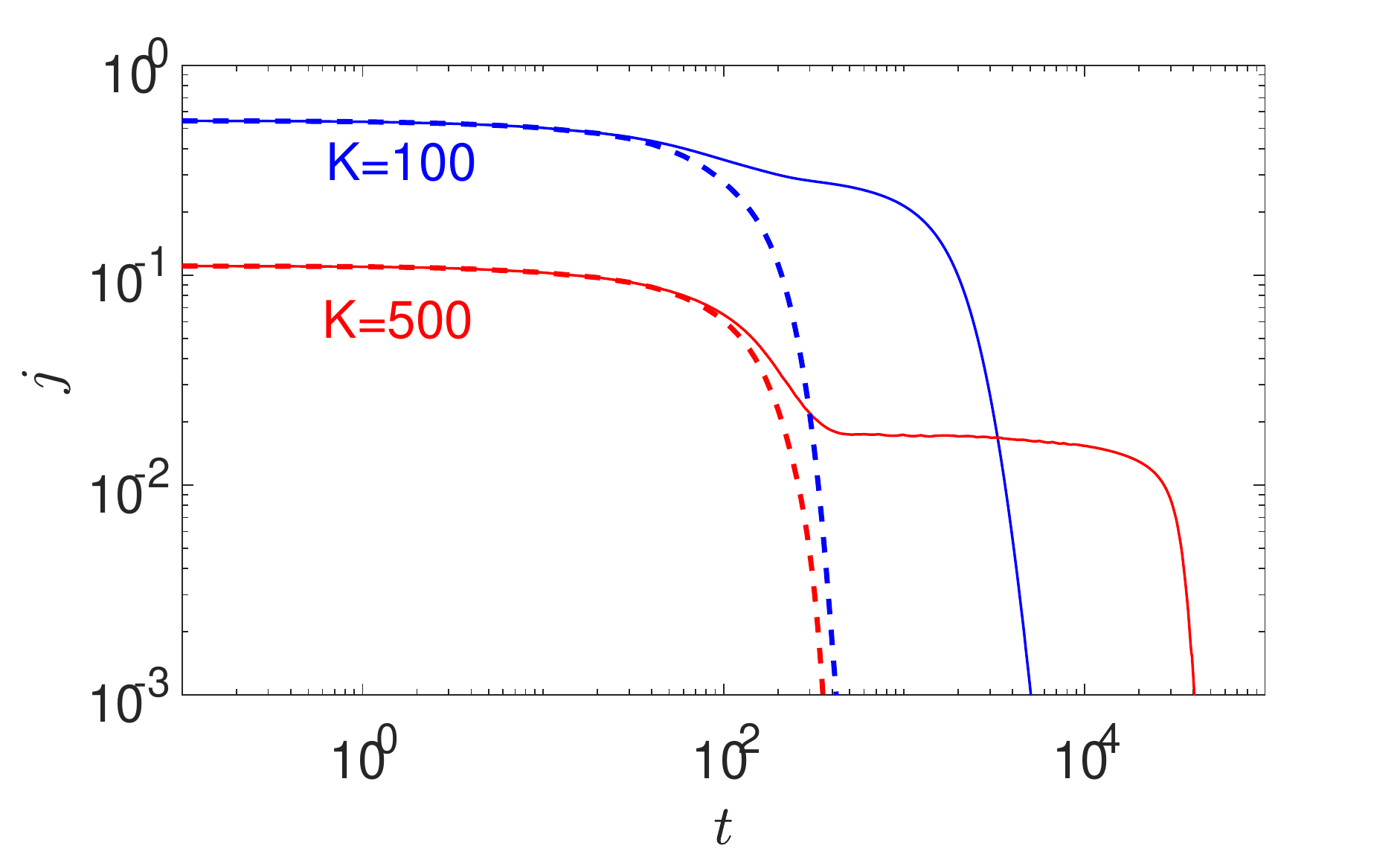}
	\caption{(a) The relaxation of the electric field in the bulk (at $x=0$) over time and (b) transient currents according to~\eqref{eq:j}. In both panels, dashed lines correspond to the results without reactions(i.e., excluding~\eqref{eq:disp_react}) and solid lines correspond to results when reactions are included. The parameters are the same as in Figure~\ref{fig:p_prof}. The generation of new charged colloids, whose migration contributes to a quasi-steady state current and a further relaxation of the electric field toward electroneutrality, is apparent for high values of $K$, as shown here for $K=500$.} \label{fig:currents}
\end{figure}

{Direct numerical integration of equations~\eqref{eq:final_form}, indicates, as expected, that the charge accumulates at high voltages near the electrode in the form of a plateau whose width grows until it stabilizes at steady state (Figure~\ref{fig:p_prof}a). In the absence of disproportionation reactions, for high values of $K$, the maximal density at the electrodes will not be attained, while the charge in the bulk will be completely depleted, as shown by the dashed lines in Figures~\ref{fig:p_prof}b and Figure~\ref{fig:p_prof}c, respectively. This indicates that there is insufficient charge to counter the electric field, and indeed, as may be seen from figure~\ref{fig:currents}, the electric field in the bulk, $E_b$, remains almost identical to its initial value when disproportionation reactions are absent. When reactions are present (solid lines), however, additional charge is generated, such that the maximal density is reached at the electrode (Figure~\ref{fig:p_prof}b), a non-zero density of charged species remains left over in the bulk (Figure~\ref{fig:p_prof}c), and the electric field in the bulk shows significant relaxation toward electroneutrality (Figure~\ref{fig:currents}a).}

{The transient current density may be now computed as well by integrating the charge flux difference over the domain~\cite{neyts2010,zhao2011diffuse,yochelis2014transition}:
	\begin{equation}\label{eq:j}
	j=\frac{S}{L}\int_{-L/2}^{L/2} |{J}^p-{J}^n|\,{\rm d}x,
	\end{equation}
	where $S$ is the area of the electrodes and is taken here to be unity.} The dependence of the currents on time is shown in Figure~\ref{fig:currents}b, where we see a fast decay in the absence of disproportionation reactions, since the only contribution to the current is the migration of the charged colloids that were initially present. When reactions are present, the newly generated charged colloids drift under the influence of the incompletely screened electric field, contributing to the quasi-steady state current that is observed at high values of $K$, as has also been described by Strubbe \textit{et al.}~\cite{neyts2006,strubbe2015characterizing}. However, asymptotically, the quasi-steady state current also decays and moreover, the temporal extent depends on the disproportionation rates, such that the quasi-steady state current are less pronounced as $K$ is decreased (see Figure~\ref{fig:currents}b).

\section{From interfacial crowding to bulk charging}

Numerical integration of~\eqref{eq:final_form} can generally provide both the temporal and spatial properties. However, in the following discussion, we wish to obtain estimates for the width of the crowding layer and the concentration of charged colloids remaining in the bulk, as well as for the range of parameters for which electroneutrality and residual steady state currents may be expected.

At steady state the fluxes vanish, $\phi(x=0)=0$, and the asymptotic bulk densities are $p(x=0)=n(x=0)=m(x=0)/K=\rho_b$, where $x\in [-L/2,L/2]$. Since all species are of equal sizes, we find
\[
p=\frac{\rho_b  e^{-\phi}}{2\rho_b\cosh\phi+1-2\rho_b}, \quad
n=\frac{\rho_b  e^{\phi}}{2\rho_b\cosh\phi+1-2\rho_b},
\]
so that Poisson's equation reads as
\begin{equation}\label{poisson}
\phi_{xx}=\frac{2\rho_b \sinh\phi}{2\rho_b\cosh\phi+1-2\rho_b}.
\end{equation}
Assuming electroneutrality in the bulk region (i.e., the electric field in the middle of the domain is completely screened), multiplying equation~\eqref{poisson} by $\phi_x$ and integrating~\cite{kilic2007steric}), we obtain 
\begin{equation}
(\phi_x)^2=2\left[\ln\left(2\rho_b\cosh\phi+1-2\rho_b\right)\right]
\end{equation}
Next, we approximate $\cosh\phi\sim \exp|\phi|$ for large $|\phi|$, which for high potential requires $|\phi|\gg 1$ ($\gg k_BT$ in real units) and $|\phi|>\ln\left({1}/{(2\rho_b)}-1\right)$, i.e., (for $\phi>1$), $\rho_b>{1}/{\left(2(e+1)\right)}$. We thus find
\begin{equation}
\phi_x\sim\sqrt{2}\sqrt{\ln( 2\rho_b)+|\phi|}.
\end{equation}
After replacing the asymptotic relation with equality, separating variables and integrating between $x$ and $L/2$ ($-L/2$), with the boundary condition $\phi(\pm L/2)=\pm V/2$, we express the potential in terms applied voltage ($V$), remaining charge concentration in the bulk ($\rho_b$) and domain size ($L$):
\begin{equation}\label{eq:phi_apprx}
\phi=\pm\frac{V}{2}-\sqrt{V+2\ln(2\rho_b)}\left(\pm\frac{L}{2}-x\right)\pm\frac{1}{2}\left(\pm\frac{L}{2}-x\right)^2.
\end{equation}
This approximation is valid until some distance $\Lambda$ from the boundaries at which $|\phi|\sim O(1)$. Due to the quadratic form of~\eqref{eq:phi_apprx}, its second derivative is one, and therefore the region of validity of the approximation overlaps the saturation regions where $p=1$ ($n=1$). 

\begin{figure}[tp] 
	(a)\resizebox{0.5\hsize}{!}{\includegraphics[width=0.5\textwidth]{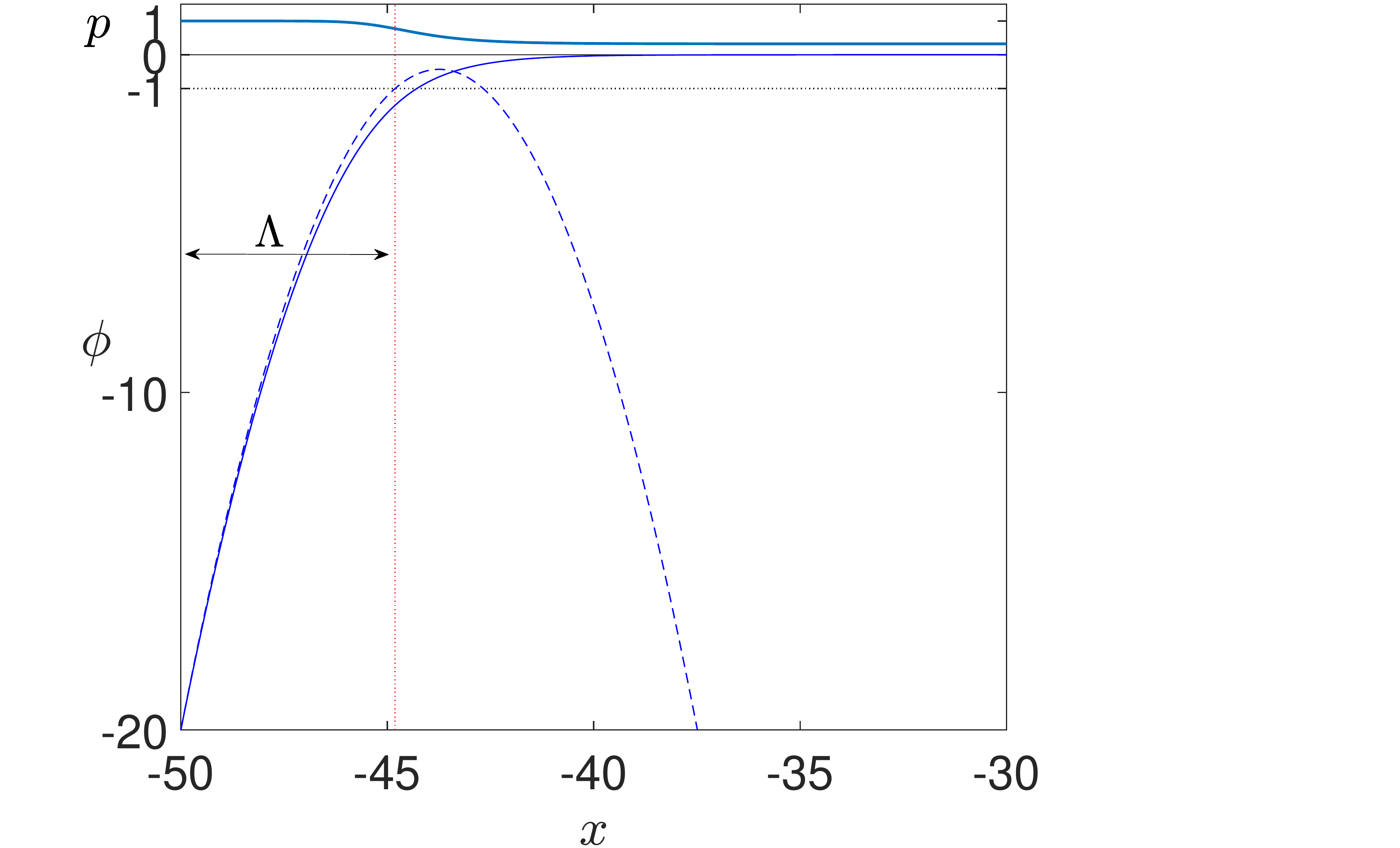}} 
	(b)\resizebox{0.5\hsize}{!}{\includegraphics[width=0.5\textwidth]{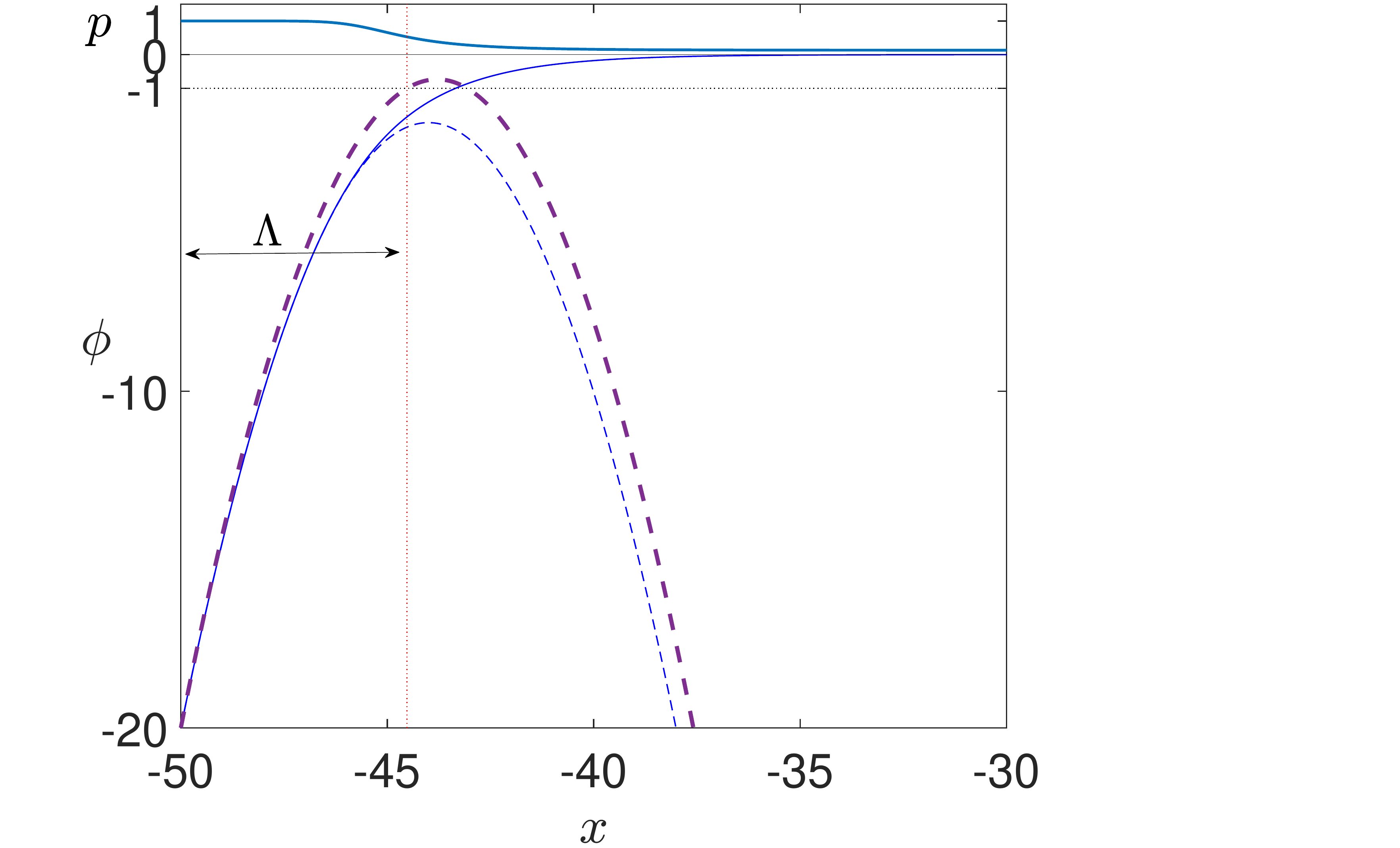}} 
	\caption{{A comparison of the approximation~\eqref{eq:phi_apprx} (bottom panels) obtained near the boundary at $x=-L/2$ for high $\phi$ (dashed thin curves) and the potential profile obtained by direct numerical integration of~\eqref{eq:final_form} (solid curve), together with the respective profiles for $p$ (top panels). The estimated saturated charge layer width, $\Lambda$, according to~\eqref{eq:delta_highphi}, is identified by the vertical dotted line. Parameters: $V=40$, $c=0.7$, $L/2=50$, and (a) $K=0.05$, (b) $K=3$. 
			For $K=3$, as in (b), $\rho_b<1/2e$ for certain values of $c$ (see Fig.~\ref{fig:p_bulk_varyingChi}), so that~\eqref{eq:phi_apprx} and thus $\Lambda$ are approximated by using~\eqref{eq:lin_approx}, which effectively stretches the parabolic form of the potential, such that the minimum/maximum falls within $\phi \in [-1,1]$ (thick dashed line).}} \label{fig:phi_approx}
\end{figure}

{$\Lambda$ therefore marks the width of the saturation plateau that forms near the boundary, as shown in Figure~\ref{fig:phi_approx}. To approximate the width of the saturation plateau, we start by solving for $\phi(\pm L/2\mp\Lambda)=\pm 1$ (i.e., comparing to $k_BT$ in real units) as a necessary condition, yielding
	\begin{equation}\label{eq:delta_highphi}
	\Lambda=\sqrt{V+2\ln(2\rho_b)}-\sqrt{2}\sqrt{\ln(2\rho_b)+1}.
	\end{equation}
	This distance depends on the unknown $\rho_b$, which in turn depends on the spontaneously generated charge in the bulk region, which will be determined immediately. Specifically, we will show that although approximation for $\Lambda$ formally requires that $\rho_b>1/2e$, it is possible to exploit a Taylor expansion to obtain a good agreement for $\rho_b<1/2e$ as well.} 

We start with conservation of the total number of colloids, namely, 
\begin{equation}
\frac{1}{L}\int_{-L/2}^{L/2}{\rm d}x \Bra{p(x,t)+n(x,t)+m(x,t)}=c,
\end{equation}
and also consider a large domain for which the width of the transition zone between the bulk and the boundary layer can be approximated as a sharp interface. The final distributions may therefore be approximated as being uniformly equal to the saturation densities or zero in the boundary layers and uniformly equal to their equilibrium bulk values ($p=n=m/K=\rho_b$) in between the boundary layers, e.g.,
\begin{equation}
p(x,t\to\infty)=
\begin{cases} 
&\rho_b, \quad -\frac{L}{2}+\Lambda<x<\frac{L}{2}-\Lambda\\
& 0, \quad \frac{L}{2}-\Lambda<x<\frac{L}{2}\\
& 1, \quad -\frac{L}{2}<x<-\frac{L}{2}+\Lambda
\end{cases},
\end{equation}
with $n$ and $m$ defined similarly.
This provides a second algebraic equation:
\begin{equation}\label{eq:delta_micellenumber}
\Lambda=\frac{L}{2}\frac{c-(2+K)\rho_b}{1-(2+K)\rho_b},
\end{equation}
which, together with \eqref{eq:delta_highphi}, yields $\Lambda$ and $\rho_b$. Solutions to these equations require that the colloid concentration and the charging fraction satisfy the relation ${1}/{2e}<\rho_b<c/(2+K)<{1}/{(2+K)}$, such that 
\[
K<2e-2\approx 3.4, 
\]
which corresponds to a significant fraction of charged colloids, i.e., about 50\% and higher. This condition appears to mark off a very limited sliver of $c$ and $K$ values for which the approximation \eqref{eq:delta_highphi} is informative. These limits may be circumvented, however, by expanding $\ln(2\rho_b)\simeq 2\rho_b-1+O(\rho_b^2)$ when $\rho_b<{1}/{2e}$, to approximate~\eqref{eq:delta_highphi} as
\begin{equation}\label{eq:series_expansion}
\Lambda \simeq \sqrt{V+2(2\rho_b-1)}-2\sqrt{2 \rho_b},
\end{equation}
where $\rho_b$ is solved through 
\begin{equation}\label{eq:root}
\sqrt{V+2(2\rho_b-1)}-2\sqrt{2 \rho_b}-\frac{L}{2}\frac{c-(2+K)\rho_b}{1-(2+K)\rho_b}=0,
\end{equation}
where the root falls in the interval $\rho_b\in [0,1/(2+K)]$.
\begin{figure}[tp] 
	\resizebox{0.5\hsize}{!}{\includegraphics{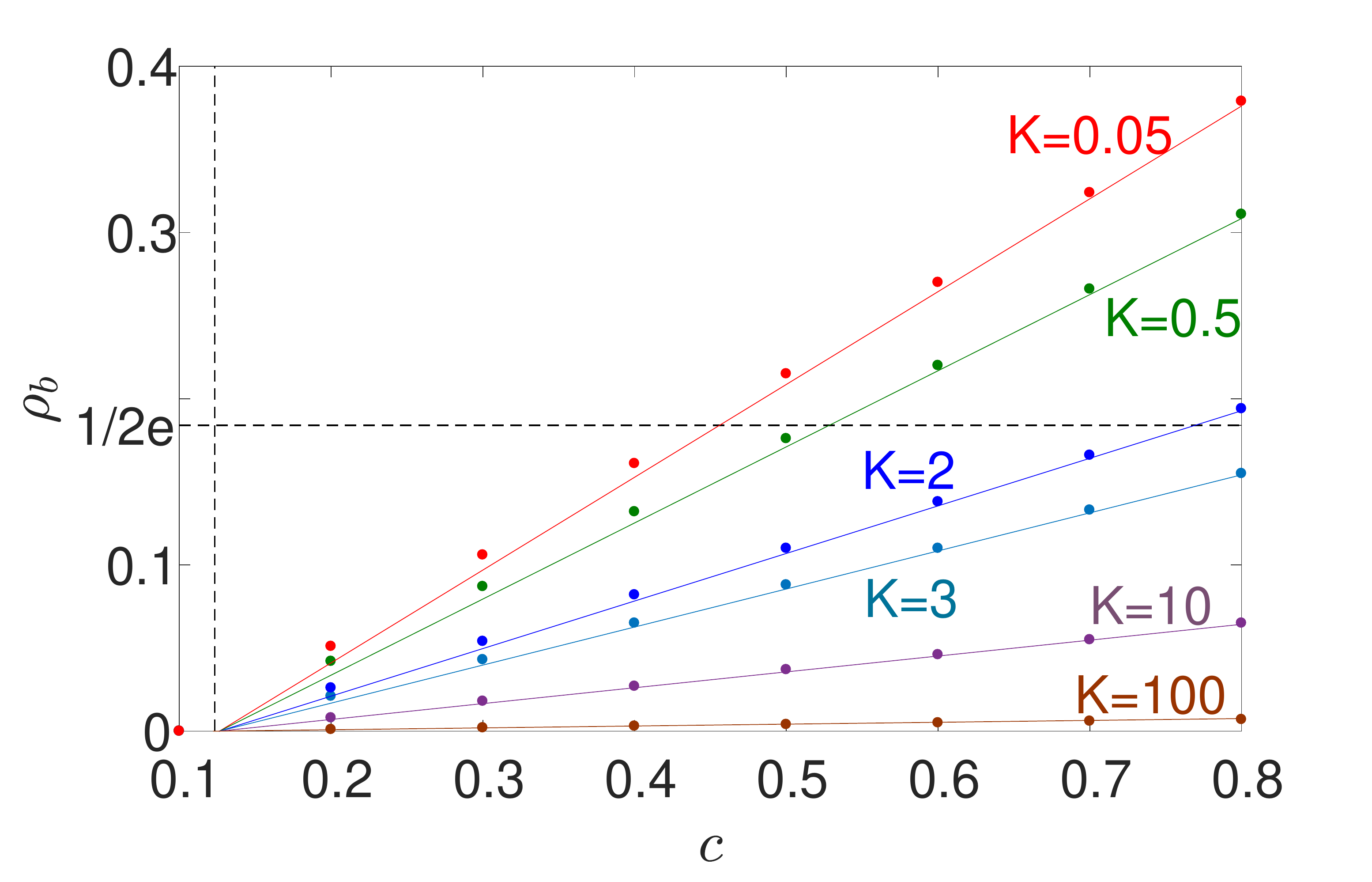}}
	\caption{{Dependence of $\rho_b$, see~\eqref{eq:lin_approx}, according to~\eqref{eq:delta_highphi} and~\eqref{eq:series_expansion} on the total concentration $c$ for $V=40$ and $L=100$. The dots show the asymptotic results of direct numerical integration of~\eqref{eq:final_form}. The vertical dashed line marks the total concentration $c_{c}$ (see~\eqref{eq:c_c}) below which the bulk charge is completely depleted.}} \label{fig:p_bulk_varyingChi}
\end{figure}
\begin{figure}[tp]
	(a)\includegraphics[width=0.5\textwidth]{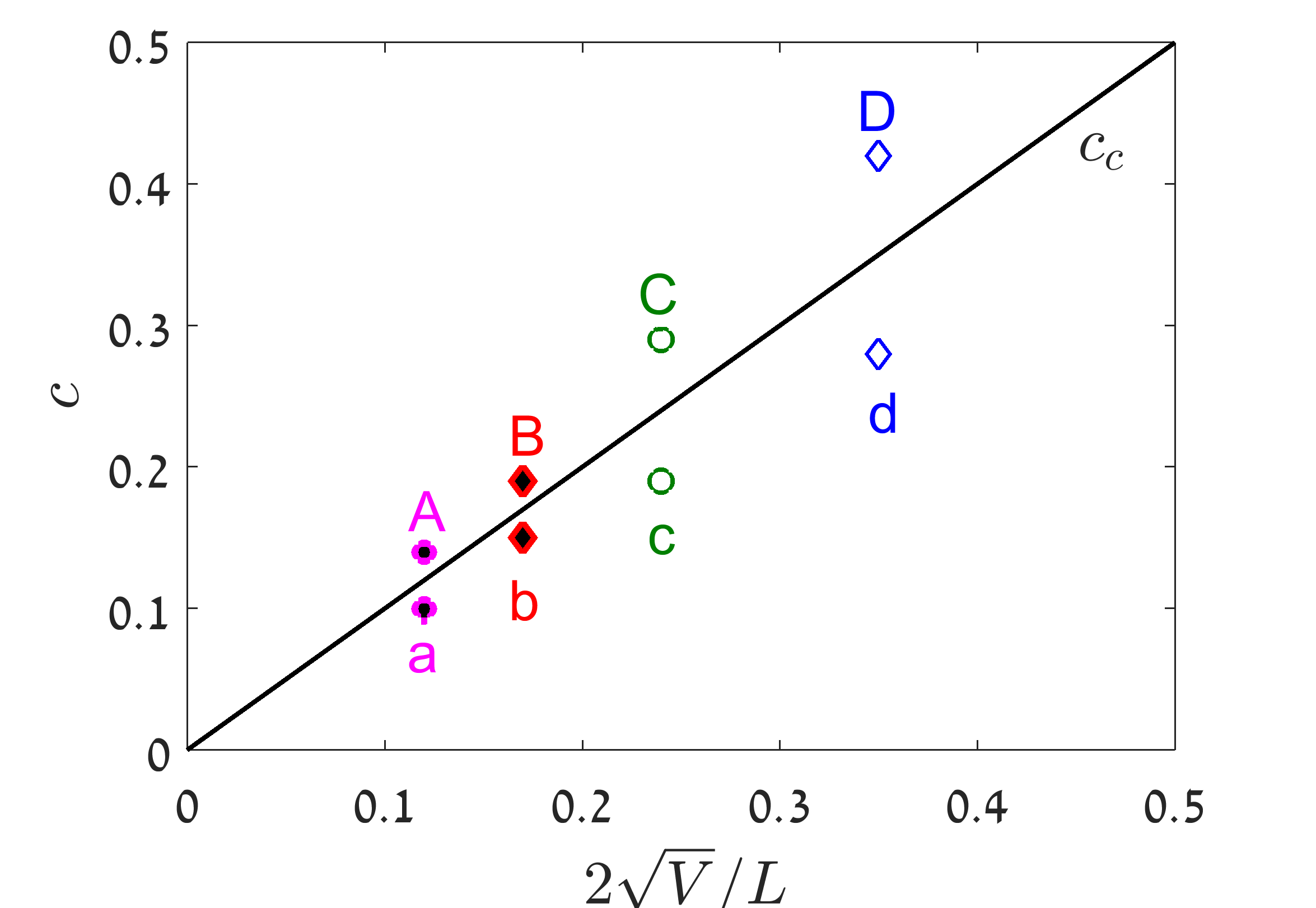}
	(b)\includegraphics[width=0.5\textwidth]{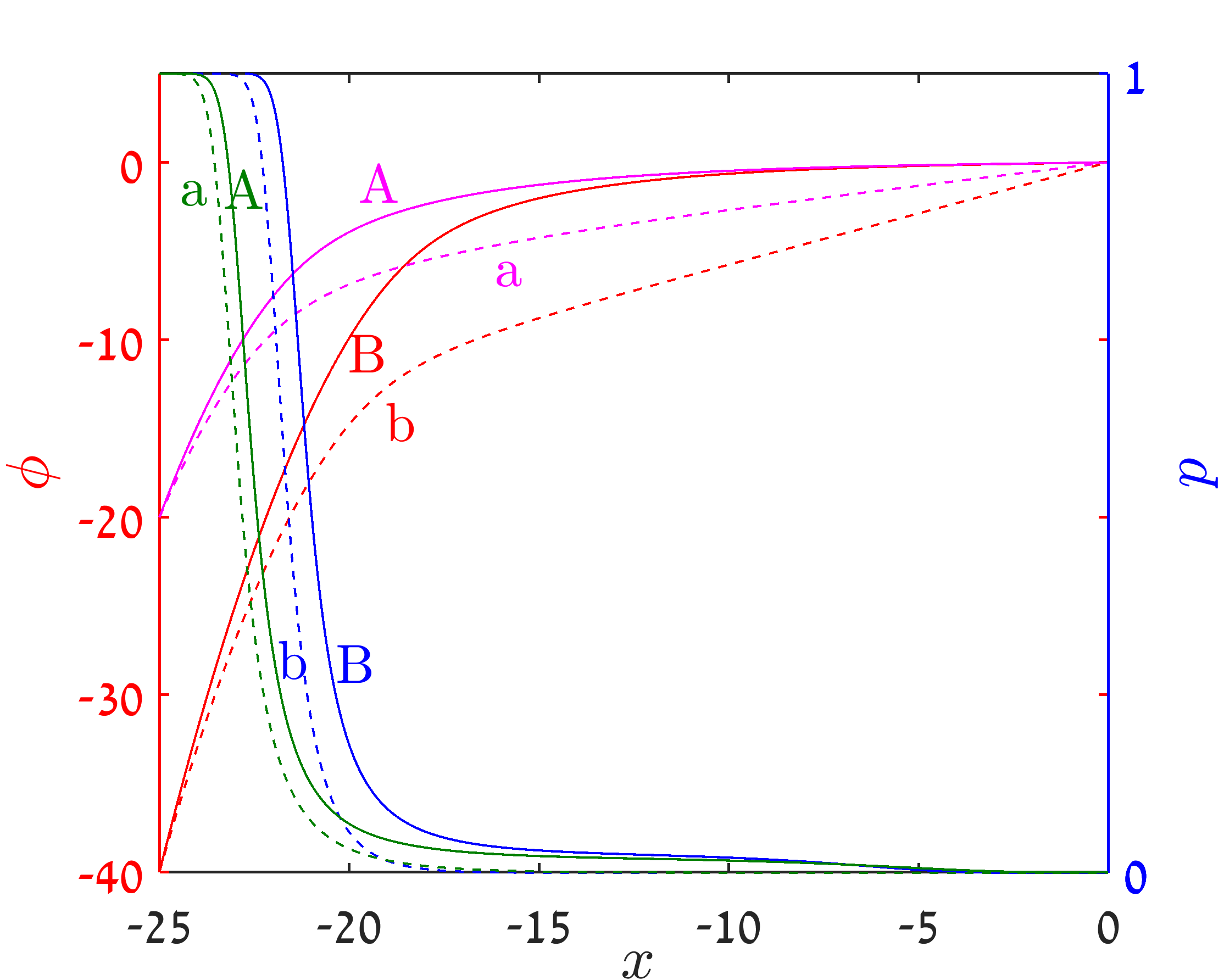} 
	(c)\includegraphics[width=0.5\textwidth]{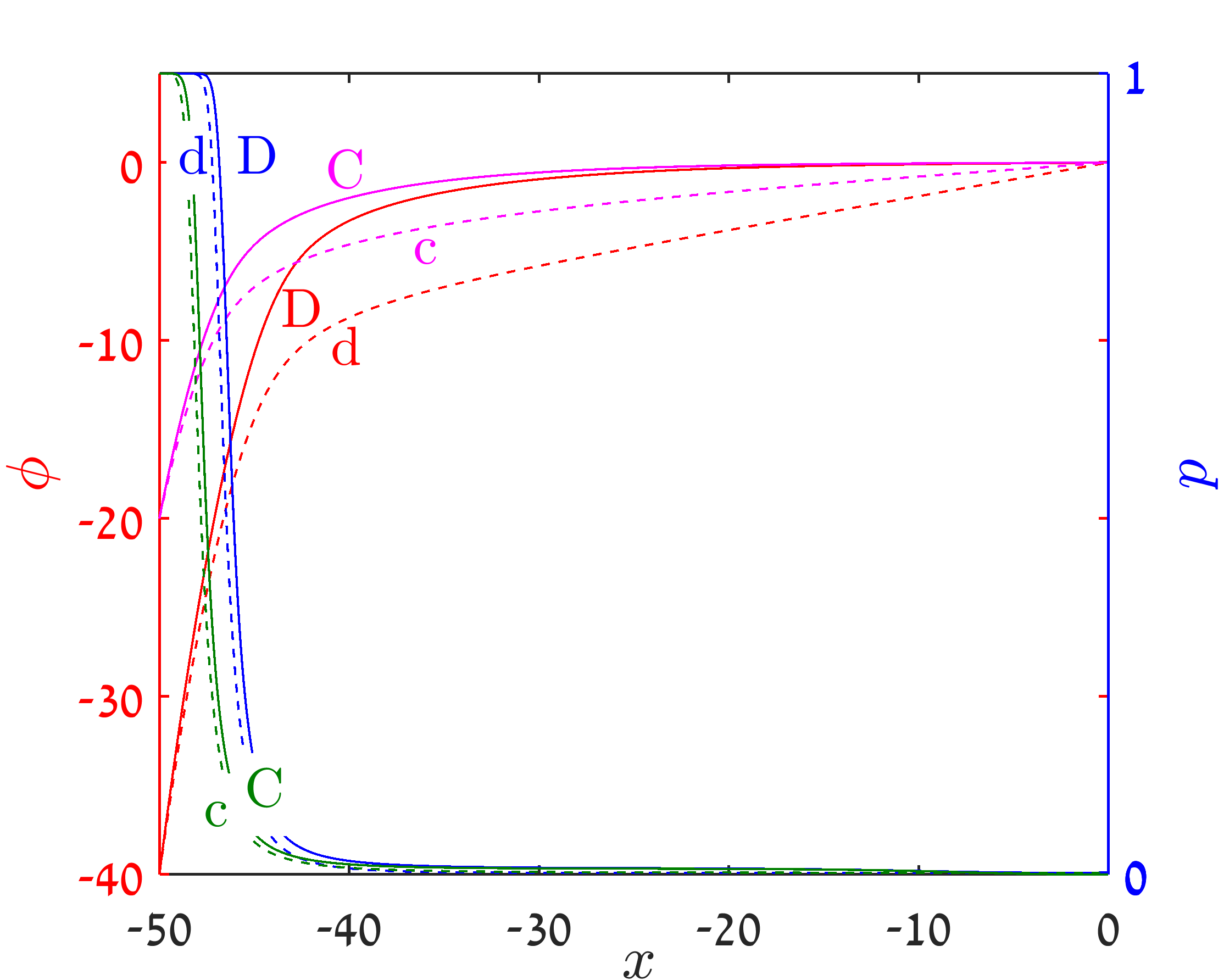} 
	\caption{{Validation of the transition from incomplete to full depletion of bulk charge and the consequential violation of electroneutrality. (a) Parameter plane which shows the critical concentration $c_c$ (solid line) according to~\eqref{eq:c_c} and deviations taken at values of $c=(1\pm 0.2)c_{c}$, for $V=40$ (circles), $V=80$ (diamonds), $L=100$ (black) and $L=50$ (white). The asymptotic profiles corresponding the values of $c$ taken in (a) for charge (right vertical axis, dark colors) and potential (left vertical axis, light colors) are shown in (b) and (c) for $L=50$ and $L=100$, respectively. The transition between the complete screening of the electric field in the bulk and the incomplete screening due to charge depletion is seen by comparing the solid lines, corresponding to $c>c_{c}$, to the dashed lines, correspond to $c<c_{c}$.}} 
	\label{fig:chi_c}
\end{figure}
Since our interest is in the case where both $V,L\gg1$, while $\rho_b<1$, the quadratic equation may be further simplified via dividing by $L$, taking the first term to be $\sqrt{V}$ and neglecting the second term. This provides an approximation to $\rho_b$ that has a linear form, as is also shown in Fig.~\ref{fig:p_bulk_varyingChi}:
\begin{equation}\label{eq:lin_approx}
\rho_b=\frac{1}{2+K}\frac{c-2\sqrt{V}/L}{1-2\sqrt{V}/L}.
\end{equation}
In the absence of charged colloids ($\rho_b \to 0$), this gives a critical concentration below which, for a given voltage and domain size, the bulk charge is depleted, and the electric field in the bulk is incompletely screened:
\begin{equation}\label{eq:c_c}
c_{c}\to\frac{\sqrt{V}}{L/2}\sim \frac{\sqrt{\rm applied~voltage}}{\rm one~half~effective~ domain~length}.
\end{equation}
The inverse statement {should be exercised with care - at large $K$ values, $\rho_b\to 0$ for $c<c_{c}$, yet the electric field will persist in the bulk even for concentrations above $c_{c}$.} In fact, this is the region in which a residual (quasi-steady state) current due to charge generation is observed, as shown in Figure~\ref{fig:currents}.

The comparison between the predicted values of $\rho_b$ and the values obtained by numerical simulations are shown in Figure~\ref{fig:p_bulk_varyingChi}, where the circles denote the simulation results. The dashed line demarcates the values of $c$ below which the series expansion ~\eqref{eq:series_expansion} must be used. The transition between a completely screened electric field and an incompletely screened field due to charge depletion is apparent when examining the profiles of the charge density $p$ and potential $\phi$ above and below $c_{c}$, as shown in Figure~\ref{fig:chi_c}. Since the accuracy of the uniform-density approximations made in deriving $c_{c}$ is worse on smaller domains, where the size of the transition zone between the saturated boundary layer and the bulk becomes more significant, the change in the behavior of the system is best visualized by examining values of $c$ at approximately 20\% above and below $c_{c}$. \\

\section{Conclusions}

We have presented a continuum model of a dispersion of monovalent and monodispersed charged and neutral colloids in a non-polar solution. The model extends the Poisson-Nernst-Planck equations to include high voltages and {colloids with a high degree of charging}. Analysis of the model at high voltages provides an estimation of the width of the charge saturation plateau that forms near the boundary due to steric crowding. This in turn enables the prediction of the bulk concentrations of the charged and neutral species, and the total colloid concentration below which the charge is depleted in the bulk, i.e., the level of dilution by solvent at which the high voltage causes all charged species to accumulate in the double layer near the electrode, with insufficient charges or charge sources (colliding neutral inverse micelles) remaining in the bulk to complete the task of screening the electric field. To summarize the results in dimensional units, we estimate the following quantities
\begin{description}[style=unboxed,leftmargin=0cm]
	\item[\rm \it residual bulk charge at equilibrium]
	\[
	\rho_b=\frac{1}{2+K}\frac{Lc-2\sqrt{\frac{\epsilon V\rho_{\rm max}}{q}}}{L-2\sqrt{\frac{\epsilon V}{q\rho_{\rm max}}}},
	\]
	{with $c$ the total colloid concentration in dimensional units,}\\
	\item[\rm \it width of charge plateau near the electrode]
	\[
	\Lambda=\frac{L}{2}\frac{c-(2+K)\rho_b}{\rho_{\rm max}-(2+K)\rho_b},
	\]
	\item[\rm \it criterion for complete bulk charge depletion]
	\[	
	c<\frac{L}{2}\frac{q}{k_BT}\sqrt{\frac{\rho_{\rm max} qV}{\epsilon}}\rho_{\rm max}.
	\]
\end{description}
We note that the validity of the above criterion is limited when the ratio of charged to neutral colloids is low ($K \gg 2$), since the electric field is not screened at steady state even when the total colloid concentration is high. This situation reflects the quasi-steady state currents observed in experiments with inverse micelles~\cite{neyts2017}.  

The scope of this study is rather phenomenological, as it ignores dependencies or interactions between the model parameters. For example, the reaction rates are assumed to be independent parameters, whereas in inverse micelles size has been shown to affect reactivity~\cite{neyts2017} (with larger inverse micelles corresponding to a greater ratio of charged to neutral inverse micelles). Nevertheless, the results provide intuitive guidelines that may be used when designing systems of mobile charged species where control over the double layer or bulk concentrations is desired, such as in the cases of e-ink devices or nano-particle synthesis. More generally, the formulated equations may also serve as a useful starting point to approach the question of the amount of free charge versus charge bound up in ion pairs or aggregates in solutions of highly concentrated electrolytes~\cite{adar2017bjerrum}, such as ionic liquid mixtures~\cite{gebbie2013ionic,lee2014room,kirchner2015ion}. The model equations and high voltage approximations may be used with different reaction mechanisms, such as the formation of one neutral macromolecule (with different size) rather than two ($p+n\rightleftharpoons \widetilde m$), although the solution for the bulk concentration ($\rho_b$) and the width of the crowded layer ($\Lambda$) is complicated by total number of colloids not being conserved in this case. The extension of these results to different and complex reaction mechanisms~\cite{molina2007electrochemistry,tonova2008reversed,vanag2008patterns} and in relation to polar media~\cite{baruah2006water,dos2016adsorption} remains open to further study. 

\begin{acknowledgements}
	The research was partially supported by the Adelis Foundation for renewable energy research.
\end{acknowledgements}


\bibliographystyle{abbrv}

\begin{thebibliography}{50}%
	\makeatletter
	\providecommand \@ifxundefined [1]{%
		\@ifx{#1\undefined}
	}%
	\providecommand \@ifnum [1]{%
		\ifnum #1\expandafter \@firstoftwo
		\else \expandafter \@secondoftwo
		\fi
	}%
	\providecommand \@ifx [1]{%
		\ifx #1\expandafter \@firstoftwo
		\else \expandafter \@secondoftwo
		\fi
	}%
	\providecommand \natexlab [1]{#1}%
	\providecommand \enquote  [1]{``#1''}%
	\providecommand \bibnamefont  [1]{#1}%
	\providecommand \bibfnamefont [1]{#1}%
	\providecommand \citenamefont [1]{#1}%
	\providecommand \href@noop [0]{\@secondoftwo}%
	\providecommand \href [0]{\begingroup \@sanitize@url \@href}%
	\providecommand \@href[1]{\@@startlink{#1}\@@href}%
	\providecommand \@@href[1]{\endgroup#1\@@endlink}%
	\providecommand \@sanitize@url [0]{\catcode `\\12\catcode `\$12\catcode
		`\&12\catcode `\#12\catcode `\^12\catcode `\_12\catcode `\%12\relax}%
	\providecommand \@@startlink[1]{}%
	\providecommand \@@endlink[0]{}%
	\providecommand \url  [0]{\begingroup\@sanitize@url \@url }%
	\providecommand \@url [1]{\endgroup\@href {#1}{\urlprefix }}%
	\providecommand \urlprefix  [0]{URL }%
	\providecommand \Eprint [0]{\href }%
	\providecommand \doibase [0]{http://dx.doi.org/}%
	\providecommand \selectlanguage [0]{\@gobble}%
	\providecommand \bibinfo  [0]{\@secondoftwo}%
	\providecommand \bibfield  [0]{\@secondoftwo}%
	\providecommand \translation [1]{[#1]}%
	\providecommand \BibitemOpen [0]{}%
	\providecommand \bibitemStop [0]{}%
	\providecommand \bibitemNoStop [0]{.\EOS\space}%
	\providecommand \EOS [0]{\spacefactor3000\relax}%
	\providecommand \BibitemShut  [1]{\csname bibitem#1\endcsname}%
	\let\auto@bib@innerbib\@empty
	\bibitem [{\citenamefont {Strubbe}\ and\ \citenamefont
		{Neyts}(2017)}]{neyts2017}%
	\BibitemOpen
	\bibfield  {author} {\bibinfo {author} {\bibfnamefont {F.}~\bibnamefont
			{Strubbe}}\ and\ \bibinfo {author} {\bibfnamefont {K.}~\bibnamefont
			{Neyts}},\ }\href@noop {} {\bibfield  {journal} {\bibinfo  {journal} {Journal
				of Physics: Condensed Matter}\ }\textbf {\bibinfo {volume} {29}},\ \bibinfo
		{pages} {453003} (\bibinfo {year} {2017})}\BibitemShut {NoStop}%
	\bibitem [{\citenamefont {Lyklema}(2013)}]{LYKLEMA2013116}%
	\BibitemOpen
	\bibfield  {author} {\bibinfo {author} {\bibfnamefont {J.}~\bibnamefont
			{Lyklema}},\ }\href@noop {} {\bibfield  {journal} {\bibinfo  {journal}
			{Current Opinion in Colloid and Interface Science}\ }\textbf {\bibinfo
			{volume} {18}},\ \bibinfo {pages} {116 } (\bibinfo {year}
		{2013})}\BibitemShut {NoStop}%
	\bibitem [{\citenamefont {Dukhin}\ and\ \citenamefont
		{Parlia}(2013)}]{DUKHIN201393}%
	\BibitemOpen
	\bibfield  {author} {\bibinfo {author} {\bibfnamefont {A.}~\bibnamefont
			{Dukhin}}\ and\ \bibinfo {author} {\bibfnamefont {S.}~\bibnamefont
			{Parlia}},\ }\href@noop {} {\bibfield  {journal} {\bibinfo  {journal}
			{Current Opinion in Colloid and Interface Science}\ }\textbf {\bibinfo
			{volume} {18}},\ \bibinfo {pages} {93 } (\bibinfo {year} {2013})}\BibitemShut
	{NoStop}%
	\bibitem [{\citenamefont {Bizmark}\ and\ \citenamefont
		{Ioannidis}(2018)}]{C8SM008D}%
	\BibitemOpen
	\bibfield  {author} {\bibinfo {author} {\bibfnamefont {N.}~\bibnamefont
			{Bizmark}}\ and\ \bibinfo {author} {\bibfnamefont {M.~A.}\ \bibnamefont
			{Ioannidis}},\ }\href@noop {} {\bibfield  {journal} {\bibinfo  {journal}
			{Soft Matter}\ }\textbf {\bibinfo {volume} {14}},\ \bibinfo {pages} {6404}
		(\bibinfo {year} {2018})}\BibitemShut {NoStop}%
	\bibitem [{\citenamefont {Smith}\ and\ \citenamefont
		{Eastoe}(2013)}]{smith2013controlling}%
	\BibitemOpen
	\bibfield  {author} {\bibinfo {author} {\bibfnamefont {G.~N.}\ \bibnamefont
			{Smith}}\ and\ \bibinfo {author} {\bibfnamefont {J.}~\bibnamefont {Eastoe}},\
	}\href@noop {} {\bibfield  {journal} {\bibinfo  {journal} {Physical Chemistry
				Chemical Physics}\ }\textbf {\bibinfo {volume} {15}},\ \bibinfo {pages} {424}
		(\bibinfo {year} {2013})}\BibitemShut {NoStop}%
	\bibitem [{\citenamefont {Prasad}\ \emph {et~al.}(2016)\citenamefont {Prasad},
		\citenamefont {Strubbe}, \citenamefont {Beunis},\ and\ \citenamefont
		{Neyts}}]{prasad2016different}%
	\BibitemOpen
	\bibfield  {author} {\bibinfo {author} {\bibfnamefont {M.}~\bibnamefont
			{Prasad}}, \bibinfo {author} {\bibfnamefont {F.}~\bibnamefont {Strubbe}},
		\bibinfo {author} {\bibfnamefont {F.}~\bibnamefont {Beunis}}, \ and\ \bibinfo
		{author} {\bibfnamefont {K.}~\bibnamefont {Neyts}},\ }\href@noop {}
	{\bibfield  {journal} {\bibinfo  {journal} {Langmuir}\ }\textbf {\bibinfo
			{volume} {32}},\ \bibinfo {pages} {5796} (\bibinfo {year}
		{2016})}\BibitemShut {NoStop}%
	\bibitem [{\citenamefont {Hsu}\ \emph {et~al.}(2005)\citenamefont {Hsu},
		\citenamefont {Dufresne},\ and\ \citenamefont {Weitz}}]{Hsu2005}%
	\BibitemOpen
	\bibfield  {author} {\bibinfo {author} {\bibfnamefont {M.~F.}\ \bibnamefont
			{Hsu}}, \bibinfo {author} {\bibfnamefont {E.~R.}\ \bibnamefont {Dufresne}}, \
		and\ \bibinfo {author} {\bibfnamefont {D.~A.}\ \bibnamefont {Weitz}},\
	}\href@noop {} {\bibfield  {journal} {\bibinfo  {journal} {Langmuir}\
		}\textbf {\bibinfo {volume} {21}},\ \bibinfo {pages} {4881} (\bibinfo {year}
		{2005})}\BibitemShut {NoStop}%
	\bibitem [{\citenamefont {Neyts}\ \emph {et~al.}(2010)\citenamefont {Neyts},
		\citenamefont {Beunis}, \citenamefont {Strubbe}, \citenamefont {Marescaux},
		\citenamefont {Verboven}, \citenamefont {Karvar},\ and\ \citenamefont
		{Verschueren}}]{neyts2010}%
	\BibitemOpen
	\bibfield  {author} {\bibinfo {author} {\bibfnamefont {K.}~\bibnamefont
			{Neyts}}, \bibinfo {author} {\bibfnamefont {F.}~\bibnamefont {Beunis}},
		\bibinfo {author} {\bibfnamefont {F.}~\bibnamefont {Strubbe}}, \bibinfo
		{author} {\bibfnamefont {M.}~\bibnamefont {Marescaux}}, \bibinfo {author}
		{\bibfnamefont {B.}~\bibnamefont {Verboven}}, \bibinfo {author}
		{\bibfnamefont {M.}~\bibnamefont {Karvar}}, \ and\ \bibinfo {author}
		{\bibfnamefont {A.}~\bibnamefont {Verschueren}},\ }\href@noop {} {\bibfield
		{journal} {\bibinfo  {journal} {Journal of Physics: Condensed Matter}\
		}\textbf {\bibinfo {volume} {22}},\ \bibinfo {pages} {494108} (\bibinfo
		{year} {2010})}\BibitemShut {NoStop}%
	\bibitem [{\citenamefont {Strubbe}\ \emph {et~al.}(2006)\citenamefont
		{Strubbe}, \citenamefont {Verschueren}, \citenamefont {Schlangen},
		\citenamefont {Beunis},\ and\ \citenamefont {Neyts}}]{neyts2006}%
	\BibitemOpen
	\bibfield  {author} {\bibinfo {author} {\bibfnamefont {F.}~\bibnamefont
			{Strubbe}}, \bibinfo {author} {\bibfnamefont {A.~R.}\ \bibnamefont
			{Verschueren}}, \bibinfo {author} {\bibfnamefont {L.~J.}\ \bibnamefont
			{Schlangen}}, \bibinfo {author} {\bibfnamefont {F.}~\bibnamefont {Beunis}}, \
		and\ \bibinfo {author} {\bibfnamefont {K.}~\bibnamefont {Neyts}},\
	}\href@noop {} {\bibfield  {journal} {\bibinfo  {journal} {Journal of Colloid
				and Interface Science}\ }\textbf {\bibinfo {volume} {300}},\ \bibinfo {pages}
		{396 } (\bibinfo {year} {2006})}\BibitemShut {NoStop}%
	\bibitem [{\citenamefont {Bikerman}(1942)}]{bikerman1942xxxix}%
	\BibitemOpen
	\bibfield  {author} {\bibinfo {author} {\bibfnamefont {J.}~\bibnamefont
			{Bikerman}},\ }\href@noop {} {\bibfield  {journal} {\bibinfo  {journal}
			{Philosophical Magazine}\ }\textbf {\bibinfo {volume} {33}},\ \bibinfo
		{pages} {384} (\bibinfo {year} {1942})}\BibitemShut {NoStop}%
	\bibitem [{\citenamefont {Borukhov}\ \emph {et~al.}(1997)\citenamefont
		{Borukhov}, \citenamefont {Andelman},\ and\ \citenamefont
		{Orland}}]{borukhov1997steric}%
	\BibitemOpen
	\bibfield  {author} {\bibinfo {author} {\bibfnamefont {I.}~\bibnamefont
			{Borukhov}}, \bibinfo {author} {\bibfnamefont {D.}~\bibnamefont {Andelman}},
		\ and\ \bibinfo {author} {\bibfnamefont {H.}~\bibnamefont {Orland}},\
	}\href@noop {} {\bibfield  {journal} {\bibinfo  {journal} {Physical Review
				Letters}\ }\textbf {\bibinfo {volume} {79}},\ \bibinfo {pages} {435}
		(\bibinfo {year} {1997})}\BibitemShut {NoStop}%
	\bibitem [{\citenamefont {Cervera}\ \emph {et~al.}(2001)\citenamefont
		{Cervera}, \citenamefont {Manzanares},\ and\ \citenamefont
		{Maf{\'e}}}]{cervera2001ion}%
	\BibitemOpen
	\bibfield  {author} {\bibinfo {author} {\bibfnamefont {J.}~\bibnamefont
			{Cervera}}, \bibinfo {author} {\bibfnamefont {J.~A.}\ \bibnamefont
			{Manzanares}}, \ and\ \bibinfo {author} {\bibfnamefont {S.}~\bibnamefont
			{Maf{\'e}}},\ }\href@noop {} {\bibfield  {journal} {\bibinfo  {journal}
			{Physical Chemistry Chemical Physics}\ }\textbf {\bibinfo {volume} {3}},\
		\bibinfo {pages} {2493} (\bibinfo {year} {2001})}\BibitemShut {NoStop}%
	\bibitem [{\citenamefont {Moreira}\ and\ \citenamefont
		{Netz}(2002)}]{moreira2002simulations}%
	\BibitemOpen
	\bibfield  {author} {\bibinfo {author} {\bibfnamefont {A.~G.}\ \bibnamefont
			{Moreira}}\ and\ \bibinfo {author} {\bibfnamefont {R.~R.}\ \bibnamefont
			{Netz}},\ }\href@noop {} {\bibfield  {journal} {\bibinfo  {journal} {European
				Physical Journal E}\ }\textbf {\bibinfo {volume} {8}},\ \bibinfo {pages} {33}
		(\bibinfo {year} {2002})}\BibitemShut {NoStop}%
	\bibitem [{\citenamefont {Cervera}\ \emph {et~al.}(2003)\citenamefont
		{Cervera}, \citenamefont {Garc{\'\i}a-Morales},\ and\ \citenamefont
		{Pellicer}}]{cervera2003ion}%
	\BibitemOpen
	\bibfield  {author} {\bibinfo {author} {\bibfnamefont {J.}~\bibnamefont
			{Cervera}}, \bibinfo {author} {\bibfnamefont {V.}~\bibnamefont
			{Garc{\'\i}a-Morales}}, \ and\ \bibinfo {author} {\bibfnamefont
			{J.}~\bibnamefont {Pellicer}},\ }\href@noop {} {\bibfield  {journal}
		{\bibinfo  {journal} {The Journal of Physical Chemistry B}\ }\textbf
		{\bibinfo {volume} {107}},\ \bibinfo {pages} {8300} (\bibinfo {year}
		{2003})}\BibitemShut {NoStop}%
	\bibitem [{\citenamefont {Kilic}\ \emph {et~al.}(2007)\citenamefont {Kilic},
		\citenamefont {Bazant},\ and\ \citenamefont {Ajdari}}]{kilic2007steric}%
	\BibitemOpen
	\bibfield  {author} {\bibinfo {author} {\bibfnamefont {M.~S.}\ \bibnamefont
			{Kilic}}, \bibinfo {author} {\bibfnamefont {M.~Z.}\ \bibnamefont {Bazant}}, \
		and\ \bibinfo {author} {\bibfnamefont {A.}~\bibnamefont {Ajdari}},\
	}\href@noop {} {\bibfield  {journal} {\bibinfo  {journal} {Physical Review
				E}\ }\textbf {\bibinfo {volume} {75}},\ \bibinfo {pages} {021503} (\bibinfo
		{year} {2007})}\BibitemShut {NoStop}%
	\bibitem [{\citenamefont {Biesheuvel}(2011)}]{biesheuvel2011two}%
	\BibitemOpen
	\bibfield  {author} {\bibinfo {author} {\bibfnamefont {P.~M.}\ \bibnamefont
			{Biesheuvel}},\ }\href@noop {} {\bibfield  {journal} {\bibinfo  {journal}
			{Journal of Colloid and Interface Science}\ }\textbf {\bibinfo {volume}
			{355}},\ \bibinfo {pages} {389} (\bibinfo {year} {2011})}\BibitemShut
	{NoStop}%
	\bibitem [{\citenamefont {Hatlo}\ \emph {et~al.}(2012)\citenamefont {Hatlo},
		\citenamefont {Van~Roij},\ and\ \citenamefont {Lue}}]{hatlo2012electric}%
	\BibitemOpen
	\bibfield  {author} {\bibinfo {author} {\bibfnamefont {M.~M.}\ \bibnamefont
			{Hatlo}}, \bibinfo {author} {\bibfnamefont {R.}~\bibnamefont {Van~Roij}}, \
		and\ \bibinfo {author} {\bibfnamefont {L.}~\bibnamefont {Lue}},\ }\href@noop
	{} {\bibfield  {journal} {\bibinfo  {journal} {EPL (Europhysics Letters)}\
		}\textbf {\bibinfo {volume} {97}},\ \bibinfo {pages} {28010} (\bibinfo {year}
		{2012})}\BibitemShut {NoStop}%
	\bibitem [{\citenamefont {Yochelis}(2014{\natexlab{a}})}]{yochelis2014spatial}%
	\BibitemOpen
	\bibfield  {author} {\bibinfo {author} {\bibfnamefont {A.}~\bibnamefont
			{Yochelis}},\ }\href@noop {} {\bibfield  {journal} {\bibinfo  {journal}
			{Journal of Physical Chemistry C}\ }\textbf {\bibinfo {volume} {118}},\
		\bibinfo {pages} {5716} (\bibinfo {year} {2014}{\natexlab{a}})}\BibitemShut
	{NoStop}%
	\bibitem [{\citenamefont {Spruijt}\ and\ \citenamefont
		{Biesheuvel}(2014)}]{spruijt2014sedimentation}%
	\BibitemOpen
	\bibfield  {author} {\bibinfo {author} {\bibfnamefont {E.}~\bibnamefont
			{Spruijt}}\ and\ \bibinfo {author} {\bibfnamefont {P.~M.}\ \bibnamefont
			{Biesheuvel}},\ }\href@noop {} {\bibfield  {journal} {\bibinfo  {journal}
			{Journal of Physics: Condensed Matter}\ }\textbf {\bibinfo {volume} {26}},\
		\bibinfo {pages} {075101} (\bibinfo {year} {2014})}\BibitemShut {NoStop}%
	\bibitem [{\citenamefont {Morrison}(1993)}]{morrison1993electrical}%
	\BibitemOpen
	\bibfield  {author} {\bibinfo {author} {\bibfnamefont {I.~D.}\ \bibnamefont
			{Morrison}},\ }\href@noop {} {\bibfield  {journal} {\bibinfo  {journal}
			{Colloids and Surfaces A: Physicochemical and Engineering Aspects}\ }\textbf
		{\bibinfo {volume} {71}},\ \bibinfo {pages} {1} (\bibinfo {year}
		{1993})}\BibitemShut {NoStop}%
	\bibitem [{\citenamefont {Comiskey}\ \emph {et~al.}(1998)\citenamefont
		{Comiskey}, \citenamefont {Albert}, \citenamefont {Yoshizawa},\ and\
		\citenamefont {Jacobson}}]{comiskey1998electrophoretic}%
	\BibitemOpen
	\bibfield  {author} {\bibinfo {author} {\bibfnamefont {B.}~\bibnamefont
			{Comiskey}}, \bibinfo {author} {\bibfnamefont {J.}~\bibnamefont {Albert}},
		\bibinfo {author} {\bibfnamefont {H.}~\bibnamefont {Yoshizawa}}, \ and\
		\bibinfo {author} {\bibfnamefont {J.}~\bibnamefont {Jacobson}},\ }\href@noop
	{} {\bibfield  {journal} {\bibinfo  {journal} {Nature}\ }\textbf {\bibinfo
			{volume} {394}},\ \bibinfo {pages} {253} (\bibinfo {year}
		{1998})}\BibitemShut {NoStop}%
	\bibitem [{\citenamefont {Mori}\ \emph {et~al.}(2001)\citenamefont {Mori},
		\citenamefont {Okastu},\ and\ \citenamefont {Tsujimoto}}]{mori2001titanium}%
	\BibitemOpen
	\bibfield  {author} {\bibinfo {author} {\bibfnamefont {Y.}~\bibnamefont
			{Mori}}, \bibinfo {author} {\bibfnamefont {Y.}~\bibnamefont {Okastu}}, \ and\
		\bibinfo {author} {\bibfnamefont {Y.}~\bibnamefont {Tsujimoto}},\ }\href@noop
	{} {\bibfield  {journal} {\bibinfo  {journal} {Journal of Nanoparticle
				Research}\ }\textbf {\bibinfo {volume} {3}},\ \bibinfo {pages} {219}
		(\bibinfo {year} {2001})}\BibitemShut {NoStop}%
	\bibitem [{\citenamefont {Grzelczak}\ \emph {et~al.}(2010)\citenamefont
		{Grzelczak}, \citenamefont {Vermant}, \citenamefont {Furst},\ and\
		\citenamefont {Liz-Marz{\'a}n}}]{grzelczak2010directed}%
	\BibitemOpen
	\bibfield  {author} {\bibinfo {author} {\bibfnamefont {M.}~\bibnamefont
			{Grzelczak}}, \bibinfo {author} {\bibfnamefont {J.}~\bibnamefont {Vermant}},
		\bibinfo {author} {\bibfnamefont {E.~M.}\ \bibnamefont {Furst}}, \ and\
		\bibinfo {author} {\bibfnamefont {L.~M.}\ \bibnamefont {Liz-Marz{\'a}n}},\
	}\href@noop {} {\bibfield  {journal} {\bibinfo  {journal} {ACS Nano}\
		}\textbf {\bibinfo {volume} {4}},\ \bibinfo {pages} {3591} (\bibinfo {year}
		{2010})}\BibitemShut {NoStop}%
	\bibitem [{\citenamefont {Zhao}\ \emph {et~al.}(2011)\citenamefont {Zhao},
		\citenamefont {He}, \citenamefont {Qiao},\ and\ \citenamefont
		{Middelberg}}]{zhao2011nanoparticle}%
	\BibitemOpen
	\bibfield  {author} {\bibinfo {author} {\bibfnamefont {C.-X.}\ \bibnamefont
			{Zhao}}, \bibinfo {author} {\bibfnamefont {L.}~\bibnamefont {He}}, \bibinfo
		{author} {\bibfnamefont {S.~Z.}\ \bibnamefont {Qiao}}, \ and\ \bibinfo
		{author} {\bibfnamefont {A.~P.}\ \bibnamefont {Middelberg}},\ }\href@noop {}
	{\bibfield  {journal} {\bibinfo  {journal} {Chemical Engineering Science}\
		}\textbf {\bibinfo {volume} {66}},\ \bibinfo {pages} {1463} (\bibinfo {year}
		{2011})}\BibitemShut {NoStop}%
	\bibitem [{\citenamefont {Yoshida}\ \emph {et~al.}(2011)\citenamefont
		{Yoshida}, \citenamefont {Kim},\ and\ \citenamefont
		{Nagaki}}]{yoshida2011green}%
	\BibitemOpen
	\bibfield  {author} {\bibinfo {author} {\bibfnamefont {J.-I.}\ \bibnamefont
			{Yoshida}}, \bibinfo {author} {\bibfnamefont {H.}~\bibnamefont {Kim}}, \ and\
		\bibinfo {author} {\bibfnamefont {A.}~\bibnamefont {Nagaki}},\ }\href@noop {}
	{\bibfield  {journal} {\bibinfo  {journal} {ChemSusChem}\ }\textbf {\bibinfo
			{volume} {4}},\ \bibinfo {pages} {331} (\bibinfo {year} {2011})}\BibitemShut
	{NoStop}%
	\bibitem [{\citenamefont {Pileni}(1993)}]{pileni1993reverse}%
	\BibitemOpen
	\bibfield  {author} {\bibinfo {author} {\bibfnamefont {M.}~\bibnamefont
			{Pileni}},\ }\href@noop {} {\bibfield  {journal} {\bibinfo  {journal}
			{Journal of Physical Chemistry}\ }\textbf {\bibinfo {volume} {97}},\ \bibinfo
		{pages} {6961} (\bibinfo {year} {1993})}\BibitemShut {NoStop}%
	\bibitem [{\citenamefont {Foudeh}\ \emph {et~al.}(2012)\citenamefont {Foudeh},
		\citenamefont {Fatanat~Didar}, \citenamefont {Veres},\ and\ \citenamefont
		{Tabrizian}}]{C2LC40630F}%
	\BibitemOpen
	\bibfield  {author} {\bibinfo {author} {\bibfnamefont {A.~M.}\ \bibnamefont
			{Foudeh}}, \bibinfo {author} {\bibfnamefont {T.}~\bibnamefont
			{Fatanat~Didar}}, \bibinfo {author} {\bibfnamefont {T.}~\bibnamefont
			{Veres}}, \ and\ \bibinfo {author} {\bibfnamefont {M.}~\bibnamefont
			{Tabrizian}},\ }\href@noop {} {\bibfield  {journal} {\bibinfo  {journal} {Lab
				Chip}\ }\textbf {\bibinfo {volume} {12}},\ \bibinfo {pages} {3249} (\bibinfo
		{year} {2012})}\BibitemShut {NoStop}%
	\bibitem [{\citenamefont {Kreutz}\ \emph {et~al.}(2010)\citenamefont {Kreutz},
		\citenamefont {Shukhaev}, \citenamefont {Du}, \citenamefont {Druskin},
		\citenamefont {Daugulis},\ and\ \citenamefont
		{Ismagilov}}]{kreutz2010evolution}%
	\BibitemOpen
	\bibfield  {author} {\bibinfo {author} {\bibfnamefont {J.~E.}\ \bibnamefont
			{Kreutz}}, \bibinfo {author} {\bibfnamefont {A.}~\bibnamefont {Shukhaev}},
		\bibinfo {author} {\bibfnamefont {W.}~\bibnamefont {Du}}, \bibinfo {author}
		{\bibfnamefont {S.}~\bibnamefont {Druskin}}, \bibinfo {author} {\bibfnamefont
			{O.}~\bibnamefont {Daugulis}}, \ and\ \bibinfo {author} {\bibfnamefont
			{R.~F.}\ \bibnamefont {Ismagilov}},\ }\href@noop {} {\bibfield  {journal}
		{\bibinfo  {journal} {Journal of the American Chemical Society}\ }\textbf
		{\bibinfo {volume} {132}},\ \bibinfo {pages} {3128} (\bibinfo {year}
		{2010})}\BibitemShut {NoStop}%
	\bibitem [{\citenamefont {Shang}\ \emph {et~al.}(2017)\citenamefont {Shang},
		\citenamefont {Cheng},\ and\ \citenamefont {Zhao}}]{shang2017emerging}%
	\BibitemOpen
	\bibfield  {author} {\bibinfo {author} {\bibfnamefont {L.}~\bibnamefont
			{Shang}}, \bibinfo {author} {\bibfnamefont {Y.}~\bibnamefont {Cheng}}, \ and\
		\bibinfo {author} {\bibfnamefont {Y.}~\bibnamefont {Zhao}},\ }\href@noop {}
	{\bibfield  {journal} {\bibinfo  {journal} {Chemical Reviews}\ }\textbf
		{\bibinfo {volume} {117}},\ \bibinfo {pages} {7964} (\bibinfo {year}
		{2017})}\BibitemShut {NoStop}%
	\bibitem [{\citenamefont {Gavish}\ \emph {et~al.}(2018)\citenamefont {Gavish},
		\citenamefont {Elad},\ and\ \citenamefont {Yochelis}}]{gavish_elad_yochelis}%
	\BibitemOpen
	\bibfield  {author} {\bibinfo {author} {\bibfnamefont {N.}~\bibnamefont
			{Gavish}}, \bibinfo {author} {\bibfnamefont {D.}~\bibnamefont {Elad}}, \ and\
		\bibinfo {author} {\bibfnamefont {A.}~\bibnamefont {Yochelis}},\ }\href@noop
	{} {\bibfield  {journal} {\bibinfo  {journal} {Journal of Physical Chemistry
				Letters}\ }\textbf {\bibinfo {volume} {9}},\ \bibinfo {pages} {36} (\bibinfo
		{year} {2018})}\BibitemShut {NoStop}%
	\bibitem [{\citenamefont {Bier}\ \emph {et~al.}(2017)\citenamefont {Bier},
		\citenamefont {Gavish}, \citenamefont {Uecker},\ and\ \citenamefont
		{Yochelis}}]{bguy}%
	\BibitemOpen
	\bibfield  {author} {\bibinfo {author} {\bibfnamefont {S.}~\bibnamefont
			{Bier}}, \bibinfo {author} {\bibfnamefont {N.}~\bibnamefont {Gavish}},
		\bibinfo {author} {\bibfnamefont {H.}~\bibnamefont {Uecker}}, \ and\ \bibinfo
		{author} {\bibfnamefont {A.}~\bibnamefont {Yochelis}},\ }\href@noop {}
	{\bibfield  {journal} {\bibinfo  {journal} {Physical Reviews E}\ }\textbf
		{\bibinfo {volume} {95}},\ \bibinfo {pages} {060201(R)} (\bibinfo {year}
		{2017})}\BibitemShut {NoStop}%
	\bibitem [{\citenamefont {Islam}(2004)}]{islam2004einstein}%
	\BibitemOpen
	\bibfield  {author} {\bibinfo {author} {\bibfnamefont {M.}~\bibnamefont
			{Islam}},\ }\href@noop {} {\bibfield  {journal} {\bibinfo  {journal} {Physica
				Scripta}\ }\textbf {\bibinfo {volume} {70}},\ \bibinfo {pages} {120}
		(\bibinfo {year} {2004})}\BibitemShut {NoStop}%
	\bibitem [{\citenamefont {Cheng}\ \emph {et~al.}(1998)\citenamefont {Cheng},
		\citenamefont {Shen}, \citenamefont {Yang}, \citenamefont {Ma}, \citenamefont
		{Tang}, \citenamefont {De~Yao},\ and\ \citenamefont
		{Sun}}]{cheng1998properties}%
	\BibitemOpen
	\bibfield  {author} {\bibinfo {author} {\bibfnamefont {G.~X.}\ \bibnamefont
			{Cheng}}, \bibinfo {author} {\bibfnamefont {F.}~\bibnamefont {Shen}},
		\bibinfo {author} {\bibfnamefont {L.~F.}\ \bibnamefont {Yang}}, \bibinfo
		{author} {\bibfnamefont {L.~R.}\ \bibnamefont {Ma}}, \bibinfo {author}
		{\bibfnamefont {Y.}~\bibnamefont {Tang}}, \bibinfo {author} {\bibfnamefont
			{K.}~\bibnamefont {De~Yao}}, \ and\ \bibinfo {author} {\bibfnamefont {P.~C.}\
			\bibnamefont {Sun}},\ }\href@noop {} {\bibfield  {journal} {\bibinfo
			{journal} {Materials Chemistry and Physics}\ }\textbf {\bibinfo {volume}
			{56}},\ \bibinfo {pages} {97} (\bibinfo {year} {1998})}\BibitemShut {NoStop}%
	\bibitem [{\citenamefont {Feldman}\ \emph {et~al.}(2002)\citenamefont
		{Feldman}, \citenamefont {Puzenko},\ and\ \citenamefont
		{Ryabov}}]{feldman2002non}%
	\BibitemOpen
	\bibfield  {author} {\bibinfo {author} {\bibfnamefont {Y.}~\bibnamefont
			{Feldman}}, \bibinfo {author} {\bibfnamefont {A.}~\bibnamefont {Puzenko}}, \
		and\ \bibinfo {author} {\bibfnamefont {Y.}~\bibnamefont {Ryabov}},\
	}\href@noop {} {\bibfield  {journal} {\bibinfo  {journal} {Chemical Physics}\
		}\textbf {\bibinfo {volume} {284}},\ \bibinfo {pages} {139} (\bibinfo {year}
		{2002})}\BibitemShut {NoStop}%
	\bibitem [{\citenamefont {Ganguly}\ and\ \citenamefont
		{Choudhury}(2012)}]{ganguly2012investigating}%
	\BibitemOpen
	\bibfield  {author} {\bibinfo {author} {\bibfnamefont {R.}~\bibnamefont
			{Ganguly}}\ and\ \bibinfo {author} {\bibfnamefont {N.}~\bibnamefont
			{Choudhury}},\ }\href@noop {} {\bibfield  {journal} {\bibinfo  {journal}
			{Journal of Colloid and Interface Science}\ }\textbf {\bibinfo {volume}
			{372}},\ \bibinfo {pages} {45} (\bibinfo {year} {2012})}\BibitemShut
	{NoStop}%
	\bibitem [{\citenamefont
		{Israelachvili}(2015)}]{israelachvili2015intermolecular}%
	\BibitemOpen
	\bibfield  {author} {\bibinfo {author} {\bibfnamefont {J.}~\bibnamefont
			{Israelachvili}},\ }\href@noop {} {\emph {\bibinfo {title} {Intermolecular
				and Surface Forces}}}\ (\bibinfo  {publisher} {Elsevier Science},\ \bibinfo
	{year} {2015})\BibitemShut {NoStop}%
	\bibitem [{\citenamefont {Gavish}\ and\ \citenamefont
		{Promislow}(2016)}]{gavish2016dependence}%
	\BibitemOpen
	\bibfield  {author} {\bibinfo {author} {\bibfnamefont {N.}~\bibnamefont
			{Gavish}}\ and\ \bibinfo {author} {\bibfnamefont {K.}~\bibnamefont
			{Promislow}},\ }\href@noop {} {\bibfield  {journal} {\bibinfo  {journal}
			{Physical Review E}\ }\textbf {\bibinfo {volume} {94}},\ \bibinfo {pages}
		{012611} (\bibinfo {year} {2016})}\BibitemShut {NoStop}%
	\bibitem [{\citenamefont {Gavish}\ \emph {et~al.}(2017)\citenamefont {Gavish},
		\citenamefont {Versano},\ and\ \citenamefont
		{Yochelis}}]{gavish2017spatially}%
	\BibitemOpen
	\bibfield  {author} {\bibinfo {author} {\bibfnamefont {N.}~\bibnamefont
			{Gavish}}, \bibinfo {author} {\bibfnamefont {I.}~\bibnamefont {Versano}}, \
		and\ \bibinfo {author} {\bibfnamefont {A.}~\bibnamefont {Yochelis}},\
	}\href@noop {} {\bibfield  {journal} {\bibinfo  {journal} {SIAM Journal on
				Applied Dynamical Systems}\ }\textbf {\bibinfo {volume} {16}},\ \bibinfo
		{pages} {1946} (\bibinfo {year} {2017})}\BibitemShut {NoStop}%
	\bibitem [{\citenamefont {Zhao}(2011)}]{zhao2011diffuse}%
	\BibitemOpen
	\bibfield  {author} {\bibinfo {author} {\bibfnamefont {H.}~\bibnamefont
			{Zhao}},\ }\href@noop {} {\bibfield  {journal} {\bibinfo  {journal} {Physical
				Review E}\ }\textbf {\bibinfo {volume} {84}},\ \bibinfo {pages} {051504}
		(\bibinfo {year} {2011})}\BibitemShut {NoStop}%
	\bibitem [{\citenamefont
		{Yochelis}(2014{\natexlab{b}})}]{yochelis2014transition}%
	\BibitemOpen
	\bibfield  {author} {\bibinfo {author} {\bibfnamefont {A.}~\bibnamefont
			{Yochelis}},\ }\href@noop {} {\bibfield  {journal} {\bibinfo  {journal}
			{Physical Chemistry Chemical Physics}\ }\textbf {\bibinfo {volume} {16}},\
		\bibinfo {pages} {2836} (\bibinfo {year} {2014}{\natexlab{b}})}\BibitemShut
	{NoStop}%
	\bibitem [{\citenamefont {Strubbe}\ \emph {et~al.}(2015)\citenamefont
		{Strubbe}, \citenamefont {Prasad},\ and\ \citenamefont
		{Beunis}}]{strubbe2015characterizing}%
	\BibitemOpen
	\bibfield  {author} {\bibinfo {author} {\bibfnamefont {F.}~\bibnamefont
			{Strubbe}}, \bibinfo {author} {\bibfnamefont {M.}~\bibnamefont {Prasad}}, \
		and\ \bibinfo {author} {\bibfnamefont {F.}~\bibnamefont {Beunis}},\
	}\href@noop {} {\bibfield  {journal} {\bibinfo  {journal} {Journal of
				Physical Chemistry B}\ }\textbf {\bibinfo {volume} {119}},\ \bibinfo {pages}
		{1957} (\bibinfo {year} {2015})}\BibitemShut {NoStop}%
	\bibitem [{\citenamefont {Adar}\ \emph {et~al.}(2017)\citenamefont {Adar},
		\citenamefont {Markovich},\ and\ \citenamefont {Andelman}}]{adar2017bjerrum}%
	\BibitemOpen
	\bibfield  {author} {\bibinfo {author} {\bibfnamefont {R.~M.}\ \bibnamefont
			{Adar}}, \bibinfo {author} {\bibfnamefont {T.}~\bibnamefont {Markovich}}, \
		and\ \bibinfo {author} {\bibfnamefont {D.}~\bibnamefont {Andelman}},\
	}\href@noop {} {\bibfield  {journal} {\bibinfo  {journal} {Journal of
				Chemical Physics}\ }\textbf {\bibinfo {volume} {146}},\ \bibinfo {pages}
		{194904} (\bibinfo {year} {2017})}\BibitemShut {NoStop}%
	\bibitem [{\citenamefont {Gebbie}\ \emph {et~al.}(2013)\citenamefont {Gebbie},
		\citenamefont {Valtiner}, \citenamefont {Banquy}, \citenamefont {Fox},
		\citenamefont {Henderson},\ and\ \citenamefont
		{Israelachvili}}]{gebbie2013ionic}%
	\BibitemOpen
	\bibfield  {author} {\bibinfo {author} {\bibfnamefont {M.~A.}\ \bibnamefont
			{Gebbie}}, \bibinfo {author} {\bibfnamefont {M.}~\bibnamefont {Valtiner}},
		\bibinfo {author} {\bibfnamefont {X.}~\bibnamefont {Banquy}}, \bibinfo
		{author} {\bibfnamefont {E.~T.}\ \bibnamefont {Fox}}, \bibinfo {author}
		{\bibfnamefont {W.~A.}\ \bibnamefont {Henderson}}, \ and\ \bibinfo {author}
		{\bibfnamefont {J.~N.}\ \bibnamefont {Israelachvili}},\ }\href@noop {}
	{\bibfield  {journal} {\bibinfo  {journal} {Proceedings of the National
				Academy of Sciences of USA}\ }\textbf {\bibinfo {volume} {110}},\ \bibinfo
		{pages} {9674} (\bibinfo {year} {2013})}\BibitemShut {NoStop}%
	\bibitem [{\citenamefont {Lee}\ \emph {et~al.}(2014)\citenamefont {Lee},
		\citenamefont {Vella}, \citenamefont {Perkin},\ and\ \citenamefont
		{Goriely}}]{lee2014room}%
	\BibitemOpen
	\bibfield  {author} {\bibinfo {author} {\bibfnamefont {A.~A.}\ \bibnamefont
			{Lee}}, \bibinfo {author} {\bibfnamefont {D.}~\bibnamefont {Vella}}, \bibinfo
		{author} {\bibfnamefont {S.}~\bibnamefont {Perkin}}, \ and\ \bibinfo {author}
		{\bibfnamefont {A.}~\bibnamefont {Goriely}},\ }\href@noop {} {\bibfield
		{journal} {\bibinfo  {journal} {Journal of Physical Chemistry Letters}\
		}\textbf {\bibinfo {volume} {6}},\ \bibinfo {pages} {159} (\bibinfo {year}
		{2014})}\BibitemShut {NoStop}%
	\bibitem [{\citenamefont {Kirchner}\ \emph {et~al.}(2015)\citenamefont
		{Kirchner}, \citenamefont {Malberg}, \citenamefont {Firaha},\ and\
		\citenamefont {Holl{\'o}czki}}]{kirchner2015ion}%
	\BibitemOpen
	\bibfield  {author} {\bibinfo {author} {\bibfnamefont {B.}~\bibnamefont
			{Kirchner}}, \bibinfo {author} {\bibfnamefont {F.}~\bibnamefont {Malberg}},
		\bibinfo {author} {\bibfnamefont {D.~S.}\ \bibnamefont {Firaha}}, \ and\
		\bibinfo {author} {\bibfnamefont {O.}~\bibnamefont {Holl{\'o}czki}},\
	}\href@noop {} {\bibfield  {journal} {\bibinfo  {journal} {Journal of
				Physics: Condensed Matter}\ }\textbf {\bibinfo {volume} {27}},\ \bibinfo
		{pages} {463002} (\bibinfo {year} {2015})}\BibitemShut {NoStop}%
	\bibitem [{\citenamefont {Molina}\ \emph {et~al.}(2007)\citenamefont {Molina},
		\citenamefont {Silber}, \citenamefont {Correa},\ and\ \citenamefont
		{Sereno}}]{molina2007electrochemistry}%
	\BibitemOpen
	\bibfield  {author} {\bibinfo {author} {\bibfnamefont {P.~G.}\ \bibnamefont
			{Molina}}, \bibinfo {author} {\bibfnamefont {J.~J.}\ \bibnamefont {Silber}},
		\bibinfo {author} {\bibfnamefont {N.~M.}\ \bibnamefont {Correa}}, \ and\
		\bibinfo {author} {\bibfnamefont {L.}~\bibnamefont {Sereno}},\ }\href@noop {}
	{\bibfield  {journal} {\bibinfo  {journal} {Journal of Physical Chemistry C}\
		}\textbf {\bibinfo {volume} {111}},\ \bibinfo {pages} {4269} (\bibinfo {year}
		{2007})}\BibitemShut {NoStop}%
	\bibitem [{\citenamefont {Tonova}\ and\ \citenamefont
		{Lazarova}(2008)}]{tonova2008reversed}%
	\BibitemOpen
	\bibfield  {author} {\bibinfo {author} {\bibfnamefont {K.}~\bibnamefont
			{Tonova}}\ and\ \bibinfo {author} {\bibfnamefont {Z.}~\bibnamefont
			{Lazarova}},\ }\href@noop {} {\bibfield  {journal} {\bibinfo  {journal}
			{Biotechnology Advances}\ }\textbf {\bibinfo {volume} {26}},\ \bibinfo
		{pages} {516} (\bibinfo {year} {2008})}\BibitemShut {NoStop}%
	\bibitem [{\citenamefont {Vanag}\ and\ \citenamefont
		{Epstein}(2008)}]{vanag2008patterns}%
	\BibitemOpen
	\bibfield  {author} {\bibinfo {author} {\bibfnamefont {V.}~\bibnamefont
			{Vanag}}\ and\ \bibinfo {author} {\bibfnamefont {I.}~\bibnamefont
			{Epstein}},\ }in\ \href@noop {} {\emph {\bibinfo {booktitle} {Self-Organized
				Morphology in Nanostructured Materials}}}\ (\bibinfo  {publisher}
	{Springer},\ \bibinfo {year} {2008})\ pp.\ \bibinfo {pages}
	{89--113}\BibitemShut {NoStop}%
	\bibitem [{\citenamefont {Baruah}\ \emph {et~al.}(2006)\citenamefont {Baruah},
		\citenamefont {Roden}, \citenamefont {Sedgwick}, \citenamefont {Correa},
		\citenamefont {Crans},\ and\ \citenamefont {Levinger}}]{baruah2006water}%
	\BibitemOpen
	\bibfield  {author} {\bibinfo {author} {\bibfnamefont {B.}~\bibnamefont
			{Baruah}}, \bibinfo {author} {\bibfnamefont {J.~M.}\ \bibnamefont {Roden}},
		\bibinfo {author} {\bibfnamefont {M.}~\bibnamefont {Sedgwick}}, \bibinfo
		{author} {\bibfnamefont {N.~M.}\ \bibnamefont {Correa}}, \bibinfo {author}
		{\bibfnamefont {D.~C.}\ \bibnamefont {Crans}}, \ and\ \bibinfo {author}
		{\bibfnamefont {N.~E.}\ \bibnamefont {Levinger}},\ }\href@noop {} {\bibfield
		{journal} {\bibinfo  {journal} {Journal of the American Chemical Society}\
		}\textbf {\bibinfo {volume} {128}},\ \bibinfo {pages} {12758} (\bibinfo
		{year} {2006})}\BibitemShut {NoStop}%
	\bibitem [{\citenamefont {dos Santos}\ \emph {et~al.}(2016)\citenamefont {dos
			Santos}, \citenamefont {Bakhshandeh}, \citenamefont {Diehl},\ and\
		\citenamefont {Levin}}]{dos2016adsorption}%
	\BibitemOpen
	\bibfield  {author} {\bibinfo {author} {\bibfnamefont {A.~P.}\ \bibnamefont
			{dos Santos}}, \bibinfo {author} {\bibfnamefont {A.}~\bibnamefont
			{Bakhshandeh}}, \bibinfo {author} {\bibfnamefont {A.}~\bibnamefont {Diehl}},
		\ and\ \bibinfo {author} {\bibfnamefont {Y.}~\bibnamefont {Levin}},\
	}\href@noop {} {\bibfield  {journal} {\bibinfo  {journal} {Soft Matter}\
		}\textbf {\bibinfo {volume} {12}},\ \bibinfo {pages} {8528} (\bibinfo {year}
		{2016})}\BibitemShut {NoStop}%
\end{thebibliography}

\end{document}